\documentclass{aastex63}

\definecolor{Red}{rgb}{1,0,0}

\received{} %June 1, 2019}
\revised{} %January 10, 2019}
\accepted{} %\today}
\submitjournal{ApJ}

\shorttitle{Excitation and Damping of Slow Magnetosonic Waves}
\shortauthors{Ofman and Wang}
\begin{document}

\title{Excitation and Damping of Slow Magnetosonic Waves in Flaring Hot Coronal Loops: Effects of Compressive Viscosity}

\correspondingauthor{Leon Ofman}
\email{ofman@cua.edu}

\author[0000-0003-0602-6693]{Leon Ofman}
\affiliation{Department of Physics\\
Catholic University of America  \\
Washington, DC 20064, USA}
\affiliation{NASA Goddard Space Flight Center \\
Code 671 \\
Greenbelt, MD 20771, USA}

\author[0000-0003-0053-1146]{Tongjiang Wang}
\affiliation{Department of Physics\\
Catholic University of America  \\
Washington, DC 20064, USA}
\affiliation{NASA Goddard Space Flight Center \\
Code 671 \\
Greenbelt, MD 20771, USA}

%\collaboration{1}{(AAS Journals Data Scientists collaboration)}



\begin{abstract}
Slow magnetosonic waves  associated with flares were observed in coronal loops by SOHO/SUMER, SDO/AIA in various EUV bandpasses, and other instruments. The excitation and damping of slow magnetosonic waves provides  information on the magnetic, temperature, and density structure of the loops. Recently, it was found using 1.5D models that the thermal conduction is suppressed and compressive viscosity is enhanced in hot ($T>6$ MK) flaring coronal loops. We model the excitation and dissipation of slow magnetosonic waves in hot coronal loops with realistic magnetic geometry, enhanced density, and temperature (compared to background corona) guided by EUV observations using 3D MHD visco-resistive model. The effects of compressive viscosity tensor component along the magnetic field are included with classical and enhanced viscosity coefficient values for the first time in 3D MHD coronal loop model. The waves are excited by a velocity pulse at the footpoint of the loop at coronal lower boundary. The modeling results demonstrate the excitation of the slow magnetosonic waves and nonlinear coupling to other wave modes, such as the kink and fast magnetosonic. We find significant leakage of the waves from the hot coronal loops with small effect of viscous dissipation in cooler (6MK) loops, and more significant effects of viscous dissipation in hotter (10.5MK) coronal loops. Our results demonstrate that nonlinear 3D MHD models are required to fully account for various wave couplings, damping, standing wave formation, and viscous dissipation in hot flaring coronal loops. Our viscous 3D MHD code provides a new tool for improved coronal seismology.

\end{abstract}



\section{Introduction} \label{intro:sec}

Observations and modeling of slow standing waves in flaring coronal loops in the context of coronal seismology, using primarily Solar and Heliospheric Observatory (SOHO) Solar Ultraviolet Measurements of Emitted Radiation
(SUMER) and Yohkoh Bragg Crystal Spectrometer (BCS) instruments observations were reviewed by \citet{Wan11}. Recent observations of slow mode waves in flaring loops were obtained by Solar Dynamics Observatory (SDO) Atmospheric Imaging Assembly (AIA) and other instruments, and analyzed in several studies \citep[e.g.,][]{Kum13,Kum15,Wan15,Man16,Pan17,Wan18,Pra21}, also, see the reviews by \citet{Wan16,Wan21}. Analysis of  slow mode observations from multiple instruments were performed and various scaling laws and the oscillations $Q$-factor (ratio of damping time to wave period) were determined by \citet{Nak19}. Recently, \citet{WO19} applied coronal seismology to SDO/AIA observations of flare-induced slow magnetosonic waves for the determination of transport coefficients, such as, thermal conduction and compressive viscosity in hot ($\sim10$ MK) coronal plasma.

The rapid (in terms of wave-periods) excitation of standing slow magnetosonic waves in coronal loops is puzzling phenomenon, since it is expected that several reflections from both footponts of the wave would be needed for the establishment of a standing wave. Various scenarios for the rapid formation of the standing wave were proposed. For example, \citet{Sel07} used 2.5D ideal MHD model to demonstrate the impulsive excitation of slow standing magnetosonic waves in a curved coronal loop due to the propagation speed differences of the fast magnetosonic pulse inside and outside the loop. The importance of 2D curvature effects on the propagation of slow magnetosonic waves was also demonstrated by \citet{GN11}. The modeling of rapid slow magnetosonic wave excitation in `cool' (1MK) coronal loops was previously extended  using 3D MHD model with similar results as the 2.5D MHD model \citep[see,][]{SO09}. In the ideal MHD models the damping of the waves was mainly due to the wave refraction out of the curved magnetic loop, wave mode coupling, and footpoint leakage.   \citet{Ofm12} studied the excitation of waves in hot coronal loops by a velocity pulse injected at the footpoint coronal boundary using 3D resistive MHD, and found that this scenario leads to generation of slow mode and coupled fast mode waves in the coronal loops, with significant wave leakage out of the loop. \citet{Fan15} studied the excitation of slow magnetosonic wave in a hot coronal loop by chromospheric evaporation flows using 2.5D MHD model. Recently,  \citet{Koh21} studied the excitation and evolution of coronal oscillations using 3D radiation MHD code Bifrost \citep{Gud11} with self-consistent simulations of solar atmosphere from the convection zone to the solar corona. However, we note that the potentially important effects of compressive viscosity on the dissipation of slow mode waves and on the rapid formation of standing waves in coronal loops were not considered in the recent studies.

The effects of compressive viscosity on the dissipation of the slow magnetosonic waves in $\sim$1 MK coronal plumes were studied in the past using 1D and 2.5D MHD models \citep{ONS00} and in coronal loops \citep{Nak00}. The damping of slow magnetosonic waves in coronal loops by thermal conduction was studied using  nonlinear 1D MHD model in the past motivated by SOHO/SUMER observations of hot flaring loops \citep{OW02}, and since then was investigated in numerous studies \cite[see reviews by][]{Wan11,Wan21}. Recently, both, thermal conduction and compressive viscosity as well as the radiative losses were considered on the damping of slow magnetosonic waves in linear and second order approximations \citep{Kum16}. Fully nonlinear 1D MHD models  of slow  magnetosonic waves that include viscosity and thermal conduction were considered \citep{Wan18,WO19,PSW21} and it was shown  that the thermal conduction must be suppressed and compressive viscosity must be enhanced compared to classical values in hot flaring loops, to account for the observed wave damping rates. A 2.5D MHD model with thermal conduction was used to study the propagation and reflection of slow magnetosonic waves in a flaring loop, such as those observed by SOHO/SUMER, SDO/AIA, and Hinode X-Ray Telescope (XRT) \citep{Fan15,Man16}. 

While most coronal and global 3D MHD models solve ideal or otherwise inviscid MHD equations, here, we use the full visco-resistive 3D MHD code NLRAT \citep{OT02,POW18,OL18}, and extend the model to include the compressive viscosity components of the viscous stress tensor \citep{Bra65}, to study for the first time the effects of compressive viscosity on the slow magnetosonic wave dissipation in realistic model of hot flaring coronal loops. Recently, \citet{POW18} used the 3D MHD code NLRAT to investigate the thin radiative cooling effects on the damping of impulsively excited slow magnetosonic waves in warm (1 MK) and hot (6 MK) coronal loops. \citet{OL18} used the 3D MHD model to study the propagation of quasi-periodic fast mode waves trains (QFPs) in bipolar active region, and found small effects due to radiative cooling and heat conduction on the fast mode magnetosonic waves.  Recently, by analyzing SDO/AIA observations of the slow magnetosonic  damping in hot flaring coronal loops it was realized that compressive viscosity may be the dominant wave damping mechanism \citep{Wan18}. Motivated by recent observational and theoretical studies, we expand the 3D MHD model with realistic magnetic geometry, density, and temperature structure of the loops to explore the role of compressive viscosity in the excitation and dissipation of slow magnetosonic waves. 

The paper is organized as follows: in Section~\ref{obs:sec} we present the observations that motivate our study, in Section~\ref{model:sec}  we present the numerical model, boundary and initial conditions, and the physical parameters, in Section~\ref{num:sec} we show the numerical modeling results. Finally, Section~\ref{disc:sec} is devoted to the discussion and conclusions.
  
\section{Observational Motivations} \label{obs:sec}

In this section we describe some observational examples of slow magnetosonic waves detected in hot flaring loops as motivation and context for our modeling study.  Longitudinal intensity oscillations detected with SDO/AIA in high-temperature EUV emission channels (94 \AA\ of 7 MK  and 131 \AA\ of 11 MK) are characterized by the flare-excited disturbances apparently traveling back and forth between the two footpoints of a hot flaring  loop system. These hot loop oscillations bearing the physical properties (e.g., periods, damping times, phase speeds, and trigger) in agreement with the Doppler shift oscillations discovered with SUMER in the flare emission lines such as Fe\,{\sc xix} and Fe\,{\sc xxi} \citep{Kli02,Wan02,Wan11} which are mainly interpreted as the fundamental standing slow mode waves \citep{OW02,Wan03a}. The intensity oscillations detected with AIA have been interpreted as either the reflected propagating slow magnetosonic waves \citep{Kum13,Kum15,nist17} or fundamental standing slow-mode waves \citep{Wan15}. A very recent study by \citet{krish21} found evidence for the transition of initial reflected propagating waves into a standing wave in the AIA 94 \AA\ channel during the loop cooling phase, confirming the theoretical predictions from 1D MHD simulations by \citet{Wan18}. The standing slow-mode oscillations were also occasionally observed in the warm (1$-$2 MK) coronal loops seen in AIA 171 \AA\ and 195 \AA\ channels \citep{Pan17}, and relatively  hot ($\sim$3 MK) loops in AIA 335 \AA\ EUV channel \citep{kim12}. 
      
\begin{table}[h]
\footnotesize
\caption{Physical Parameters of several longitudinal intensity oscillations observed with SDO/AIA.\tablenotemark{a}}
\label{wavepars:tab}
\hspace{-2cm}\begin{tabular}{cccccccccccl}
\hline\hline
Event & Time & Flare & $\lambda$ & P & $\tau$ & $A_m$ & $N_0$ & $T_0$ & $w$ & $L$ & Reference \\
 & UT & Class & \AA & min & min & & 10$^9$ cm$^{-3}$ & MK & Mm & Mm & \\
 \hline
1 & 2012/05/07 17:26 & C7.4 & 131 & 10.5 & 7.3 & 1.42 & 8.5 & 8 & 13 & 160 & \citet{Kum13} \\
2 & 2013/07/20 03:38 & C2.1 & 94 & 6.8 & 18.7 & 0.27 &  & 6.4 &  & 115 & \citet{Kum15}\\
3 & 2013/12/28 12:47 & C3.0 & 131 & 12.4 & 13.9 & 0.69 &  &  &  &  & This study\tablenotemark{c}\\
 & -- & -- & $n$\tablenotemark{b} & 12.4 & 10.7 & 0.23 & 2.6 & 8.7 & 14 & 179 & \citet{Wan15}\\
 & -- & -- & $T$\tablenotemark{b} & 11.6 & 7.7& 0.16 & -- & -- & -- &--& -- \\
4 & 2015/01/24 12:00 & B7.9 & 94 & 10 &  & & 4 & 8 & & 127 & \citet{nist17}\\
  & -- & -- & 94 & 10 & 7 & 1.5 &  &  & &  & This study\tablenotemark{c} \\
\hline
\end{tabular}
\tablenotetext{a}{Columns 2 and 3 indicate the GOES flare peak time and class. The wavelength $\lambda$ indicated the AIA imaging channels used for measuring the oscillations. The quantities $P$, $\tau$, and $A_m$ are the wave period, damping time, and relative maximum amplitude. $N_0$ and $T_0$ are the mean electron density and temperature of the loop, $w$ and $L$ are the loop width and length. The ``--" represents same value as the above row, while the space indicates no data. }
\tablenotetext{b}{$n$ and $T$ indicate that the wave parameters are measured from the oscillations in density and temperature, respectively. }
\tablenotetext{c}{The parameters are measured from the best fits to the intensity oscillations normalized to the background trend.}
\end{table}
In Table~\ref{wavepars:tab} we summarize the physical parameters of the flaring loop oscillations measured from four events reported in the literature. The wave properties such as period ($P$), damping time ($\tau$), and relative maximum amplitude ($A_m=\Delta{I_m}/I_0$) are mainly estimated from the intensity oscillations observed in the AIA 94 \AA\ or 131 \AA\ channel. For these events $P=9.9\pm2.3$ min in the range of 6.8$-$12 min, $\tau=11.7\pm5.6$ min in the range of 7$-$18.7 min, and $A_m=0.97\pm0.60$ in the range of 0.27$-$1.5. We find that the quality factor $\tau/P=1.2$ is in agreement with that for the SUMER oscillations \citep{Wan03b}. We note that the maximum amplitude of intensity perturbations relative to the loop background emission can be very large (up to $A_m=1.5$), suggesting the importance of nonlinearity in the initial stages of the damped oscillations. The large relative amplitude of the oscillations in observed emissions could be plausibly explained by the rapid cooling of the flaring loop \citep{real19}. The loop thermal properties were measured from AIA EUV images using differential emission measure (DEM) analysis (e.g., for events 1$-$3) or using Hinode/XRT data  with the filter ratio method (for event 4). The loop mean temperature is found to be $T=7.8\pm1.0$ MK and the mean density $n=(5.0\pm3.1)\times10^9$ cm$^{-3}$. The trigger of all four events listed in Table~\ref{wavepars:tab} is found to be associated with small flares at the loop's footpoint. Cool plasma ejections of velocities of several hundreds km~s$^{-1}$ were often detected near the flare site in AIA 304 \AA\ or 171 \AA\ channel \citep{Kum13,Kum15,nist17}. In contrast, the SUMER oscillations were often initiated with the hot plasma injection of velocities on the order of 100$-$300 km~s$^{-1}$, detected as a highly Doppler-shifted component of the Fe\,{\sc xix} spectra \citep{Wan15}. 

%1
\begin{figure}[ht]
\centerline{
\includegraphics[width=1.0\linewidth]{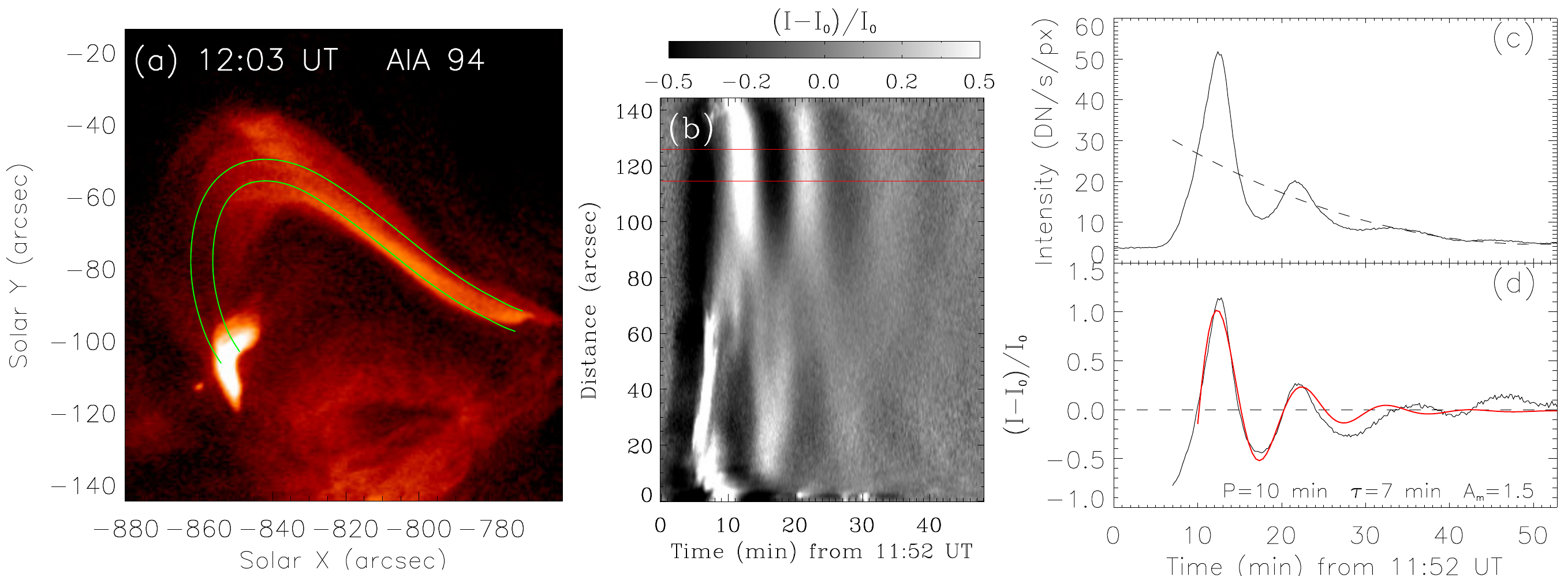}}
\caption{Longitudinal loop oscillation event observed with SDO/AIA on 24 January 2015. (a) AIA 94 \AA\ image showing the flare location and associated hot loop system. (b) Time-distance plot of intensities for a sampled slice (with average over its narrow width) along the loop (yellow strip in panel (a)). The distance is measured from the end of the slice close to the flare site. (c) Time profile of the averaged emission for a region marked in (b). The dashed curve indicates a parabolic trend. (d) Detrended time profile showing the intensity oscillations normalized to the background trend with the best fits (red solid curve). }
\label{event4:fig}
\end{figure}

%2
\begin{figure}[ht]
\centerline{
\includegraphics[width=1.0\linewidth]{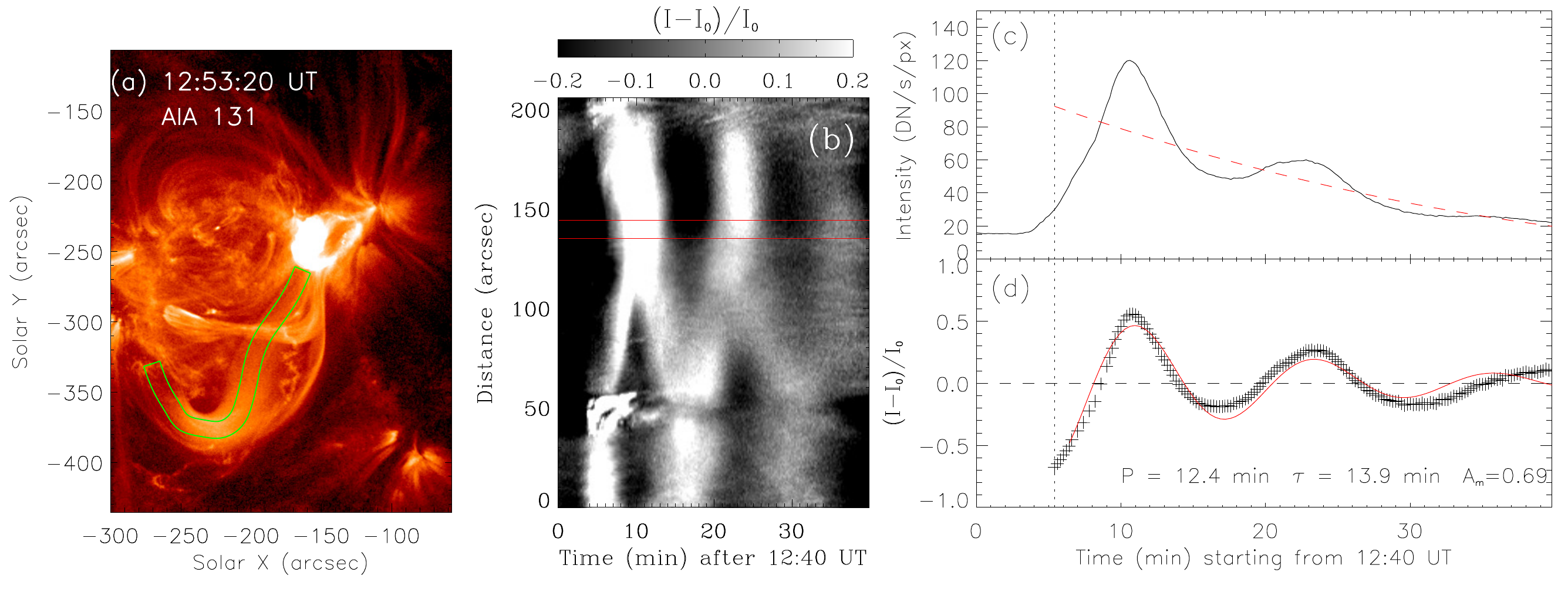}}
\caption{Longitudinal loop oscillations observed with SDO/AIA on 28 December 2013. (a) AIA 131 \AA\ image showing that a large, hot oscillatory loop was produced by a circular-ribbon flare at its footpoint. (b)-(d) Same as for Figure~\ref{event4:fig}. }
\label{event3:fig}
\end{figure}

Two examples of such longitudinal loop oscillations were illustrated here. Figure~\ref{event4:fig} shows an event occurring in the active region NOAA 12268, associated with a GOES B7.9-class flare that peaked in the soft X-ray flux at 12:00 UT on 24 January 2015 (event \#4 in Table~\ref{wavepars:tab}). This event was first analyzed by \citet{nist17} using observations from SDO/AIA, Hinode/EIS, and XRT. They found that a failed cool plasma eruption triggered simultaneous transverse kink oscillations of warm loops seen in AIA 171 \AA\ and longitudinal oscillations in a nearly co-spatial hot loop system seen in AIA 94 \AA\ and XRT Be$\_$thin filter. Note that there was an error in \citet{nist17} analysis of slow-mode oscillations (see Fig. 9 in their study) where an improper fitting of wave perturbations does not provide the correct wave amplitude and damping time, although, the measured wave  period was valid. The time-distance plot reveals the initial reflective feature and the formation of a standing wave after about two reflections as evident from the perturbations in opposite phases between the two legs (see Fig.~\ref{event4:fig}b). From the time profile of the AIA 94 \AA\ emission for the region marked in Figure~\ref{event4:fig}b in the coronal loop, we determined the oscillation parameters by best fitting to an exponentially-damped sine-function (see Fig.~\ref{event4:fig}d). We obtain $P=10.1\pm1.0$ min, $\tau=7.1\pm3.7$ min, and $A_m=1.5\pm0.6$, indicated in Table~\ref{wavepars:tab}.

Figure~\ref{event3:fig} shows another event occurring in AR 11936, associated with a GOES C 3.0-class flare that peaked at 12:47 UT on 28 December 2013 (event \#3 in Table~\ref{wavepars:tab}). This event was analyzed in detail by \citet{Wan15}. The time distance plot indicates that a fundamental standing slow magnetosonic wave has formed quickly after only one reflection of the  initial propagating disturbance (Fig.~\ref{event3:fig}b). The rapid formation  of standing waves similar to this case was previously detected by SUMER \citep[e.g.,][]{Wan03a}. We determined the oscillation parameters from the time profile of relative intensity perturbations normalized to the background trend by the best fitting of exponentially damped sinusoidal (Fig.~\ref{event3:fig}d). We obtained $P=12.4\pm1.5$ min, $\tau=13.9\pm12.8$ min, and $A_m=0.69\pm0.50$, indicated in Table~\ref{wavepars:tab}.

Using the DEM technique, \citet{Wan15} derived the time evolution of density and temperature of hot plasma in the oscillating loop for event \#3 and found that the density and temperature oscillations are nearly in phase (which is unexpected for slow magnetosonic waves in the hot coronal loop where high thermal conductivity dissipation would lead to a large phase shift) and the deduced polytropic index from their wave signatures is close to the adiabatic index of $\gamma=5/3$. The small phase shift and the value of the adiabatic index suggest the suppression of thermal conduction and therefore significant enhancement of compressive viscosity is needed to account for the rapid wave dissipation \citep[see, e.g.,][]{Wan15}. The study of \citet{Wan15} provides a new method to determine the transport coefficients of flaring plasma from evolution of their thermal properties by coronal seismology of impulsively-generated slow-mode waves. \citet{WO19} extended this work by refining the deduced values of transport coefficients using numerical parametric studies. Moreover, using 1D nonlinear MHD simulations, \citet{Wan18} found that the model with the seismology-determined transport coefficients can self-consistently produce the standing slow-mode wave as quickly (within one period) as observed (e.g., event \#3), whereas the model with the classical transport coefficients produces initially propagating slow-mode waves that require many footpoint reflections to form a standing wave \citep[e.g., event \#1; see][]{Kum13,krish21}. The quick formation of a standing wave in the model with the modified transport coefficients can be explained by the scenario that the anomalous viscosity enhancement facilitates the dissipation of higher harmonic components in the initial perturbation pulse, so that the  fundamental standing mode could quickly be produced. Thus, the competition between thermal conduction and compressive viscosity in wave damping, dependent on the transport coefficients (e.g., in the normal or anomalous conditions), may provide a unified picture accounting for excitation of a reflected propagating or standing slow magnetosonic wave as well as different transition times from the propagating mode to the standing mode as observed with SUMER and SDO/AIA. We note that previous 1D models could not account fully for wave leakage and the loop curvature effects. Therefore, in the present study we extend the model to 3D MHD that provides more realistic description of these processes.

We note that the trigger of SUMER and AIA oscillations manifests some differences in the preexisting conditions of the oscillatory loops. The SUMER oscillations are often recurrent  and the loop is hot (above $\sim$6 MK) prior to the trigger of oscillations \citep[e.g.][]{Wan03a,wan07}. Whereas all the  oscillations observed by SDO/AIA are triggered in impulsively heated hot loops that are not previously seen in high temperature AIA channels (such as 94 \AA\ and 131 \AA). However, some observational evidence suggests that the coronal loops appear to be heated likely by energetic particles or thermal front slightly earlier (on timescale of several minutes) than the observed propagation of intensity perturbations along the loops  \citep[e.g.,][]{Kum13,Wan18}. To compare the excitation of slow magnetosonic waves for the different initial loop conditions, we set up the loop models below with two sets of thermal properties: in first case a loop with the maximum temperature of 6 MK situated in the 'warm' background corona at $T_0$=2 MK, and in the second case a loop with the maximum temperature of 10.5 MK situated in the hot atmosphere of $T_0$=7 MK, guided by observations.

\section{Numerical Model, Boundary, and Initial Conditions} \label{model:sec}
We use the 3D MHD model described in \citet{OT02,POW18,OL18} with the addition of compressive viscosity terms to the 3D MHD equations based on the formulation by \citet{Bra65}, (see, also past 2.5D MHD studies of compressive dissipation of slow magnetosonic waves in coronal plumes by \citet{OND99,ONS00}). The resistive 3D MHD equations with gravity, viscous stress tensor $\Pi$, and with standard notation for the variables are
\begin{eqnarray}
&&\frac{\partial\rho}{\partial t}+\nabla\cdot(\rho\mbox{\bf V})=0,\label{cont:eq}\\
&&\frac{\partial(\rho\mbox{\bf V})}{\partial t}+\nabla\cdot\left[\rho\mbox{\bf V}\mbox{\bf V}+\left(E_up+\frac{\mbox{\bf B}\cdot \mbox{\bf B}}{2}\right)\mbox{\bf I}-\mbox{\bf BB}\right]=-\frac{1}{F_r}\rho\mbox{\bf F}_g-\nabla\cdot\Pi,\label{mom:eq}\\
&&\frac{\partial\mbox{\bf B}}{\partial t}=\nabla\times(\mbox{\bf V}\times\mbox{\bf B})+\frac{1}{S}\nabla^2\mbox{\bf B},\label{ind:eq}\\
&&\frac{\partial(\rho E)}{\partial t}+\nabla\cdot\left[\mbox{\bf V}\left(\rho E+E_up+\frac{\mbox{\bf B}\cdot \mbox{\bf B}}{2}\right)-\mbox{\bf B}(\mbox{\bf B}\cdot\mbox{\bf V})+\frac{1}{S}\nabla\times\mbox{\bf B}\times\mbox{\bf B}+\Pi\cdot\mbox{\bf V}\right] \nonumber\\	%=\nonumber\\
&&=-\frac{1}{F_r}\rho\mbox{\bf F}_g\cdot\mbox{\bf V},\label{ener:eq} 
\end{eqnarray}
where $\mbox{\bf F}_g=\frac{L_0^2}{(R_s+z-z_{min})^2}\mbox{\bf\^z}$ is the gravitational force modeled with the assumption of small height of the active region compared to the solar radius $R_s$, where $L_0=0.1R_s$, $z_{min}$ is the height of the lower boundary in the model, and $\rho E=\frac{E_up}{(\gamma-1)}+\frac{\rho V^2}{2}+\frac{B^2}{2}$ is the total energy density. The details of the normalization of the variables are described in \citet{OL18}.

%The viscous force  $\mbox{\bf F}_v=-\nabla\cdot \Pi$ is added to the right-hand side of the momentum equation {\bf (Equation~\ref{mom:eq})}, where $\Pi$ is the viscous stress tensor. 
Keeping only the terms proportional to the compressive viscosity coefficient $\eta_0$ we get the components of the viscous stress tensor
\begin{eqnarray}
&&\Pi_{mn}=-\eta_0W_{0mn}\\
&&W_{0mn}=\frac{3}{2}(h_mh_n-\frac{1}{3}\delta_{mn})(h_kh_l-\frac{1}{3}\delta_{kl})W_{kl}\\
&&W_{mn}=\frac{\partial V_m}{\partial x_n}+\frac{\partial V_n}{\partial x_m}-\frac{2}{3}\delta_{mn}\nabla\cdot\mbox{\bf V},
%\textcolor{red}{\rm I~suggest~ to~ simplify~ Eqs.(2) and (3) as \nonumber}\\
%\textcolor{red}{W_{0mn}=\frac{3}{2}(h_mh_n-\frac{1}{3}\delta_{mn})W_0} \nonumber\\
%\textcolor{red}{W_0=(h_kh_l-\frac{1}{3}\delta_{kl})W_{kl} =h_kh_l\left(\frac{\partial V_k}{\partial x_l}+\frac{\partial V_l}{\partial x_k}\right)-\frac{2}{3}\nabla\cdot\mbox{\bf V} \nonubmer}
\label{Pi:eq}
\end{eqnarray} where $h_m=B_m/B$ are the unit vectors along the magnetic field, $\delta_{mn}$ is the Kronecker delta, and the subscripts are over the coordinates (Cartesian in our model) with implicit (Einstein) summation notation. The proton compressive viscosity coefficient is given by
\begin{eqnarray}
&&\eta_0=0.96n_pT_p\tau_p, 
\label{eta0:eq}
\end{eqnarray} where $n_p$ is the proton density, $T_p$ is the proton temperature, and $\tau_p=3m_p^{1/2}T_p^{3/2}/(4\pi^{1/2}\lambda e^4n_p)$ is the proton collision time \citep{Bra65,NRL2013}, where $m_p$ is the proton mass, $e$ is the electron charge, $n_p$ is the proton density, and $\lambda$ is the Coulomb logarithm. In the present single-fluid MHD study, $T_p=T_e=T$, and $n_p=n_e=n$. In the model we treat $\eta_0$ as free parameter, since non-collisional kinetic processes such as wave-particle interaction may produce an enhanced viscosity \citep[e.g.,][]{Wan18,WO19} and study cases with `classical' (i.e., given by Equation~\ref{eta0:eq}) and `enhanced' (i.e., an order of magnitude increased $\eta_0$, and keeping the temperature dependence $T^{5/2}$) viscosity coefficients. The volumetric heating rate due to the compressive viscous dissipation can be calculated from
\begin{eqnarray}
&&S_v=-\Pi_{mn}\frac{\partial V_m}{\partial x_n},
\end{eqnarray} using implicit summation. The divergence of the viscous heating energy flux term $\nabla\cdot(\Pi\cdot\mbox{\bf V})\equiv\frac{\partial}{\partial x_n}\Pi_{mn} V_m$ is included in the  energy equation (Equation~\ref{ener:eq}) equivalent to Equation~(6.33) of \citet{Bra65}. However, due to the small energy flux of the slow magnetosonic waves, the additional heating of hot flaring loops due to the viscous dissipation of these waves is found to be negligible by comparing results with and without viscous terms in the energy equation. We note that our 3D MHD model has the capability to compute radiative cooling and thermal conduction dissipation terms, in addition to compressive viscosity. However, motivated by past modeling and observations we focus on the effects of only compressive viscosity by turning off thermal conduction and radiative losses in the present modeling study.

%3
\begin{figure}[ht]
\centerline{
\includegraphics[width=0.5\linewidth]{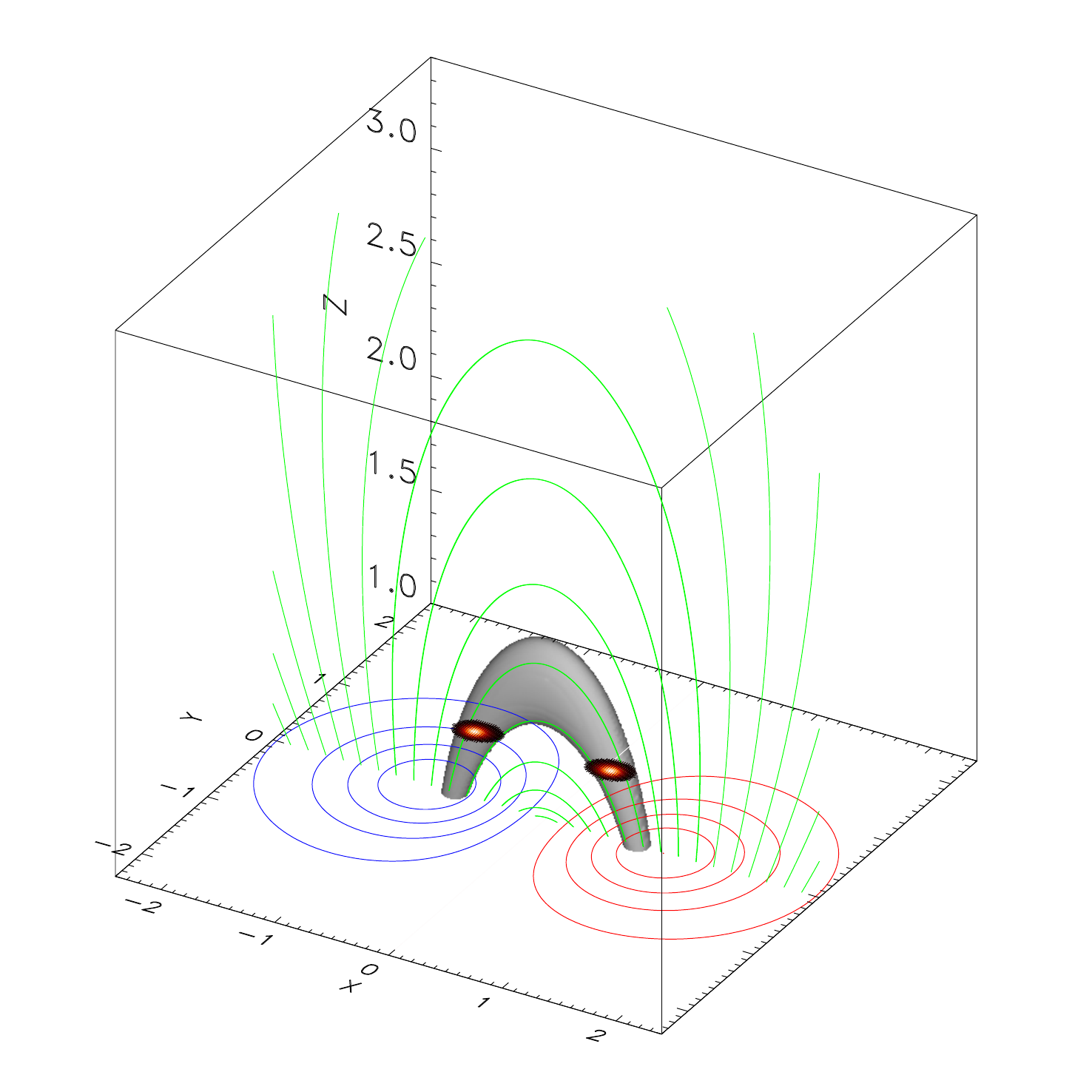}}
\caption{The initial 3D bipolar AR model with a loop with enhanced density with Gaussian cross-section distribution 3D loop model showing a dense loop in a bipolar magnetic configuration. The  isosurface for density contrast is at 1.01. The cross sections at $z=1.3$ shows the density contrast distribution in a range of 1.0 to 1.5. The contours show the z-component of magnetic fields at $z=1.0$ with levels $=\pm18,\pm36,\pm55, \pm73,\pm91$ G, where the red/blue represents the positive/negative polarity, respectively, and green indicates selected field lines in the loops' axial plane.}
\label{3Dinit:fig}
\end{figure}
The initial state and the boundary conditions of the 3D MHD model active region (AR) are described in detail in \citet{OL18}. The initial state of the coronal loop in the 3D MHD model AR is shown in Figure~\ref{3Dinit:fig}. The gray scale shows the loop density isocontour and green curves indicate several representative magnetic field lines in the plane $y=0$, and the blue and red indicate the magnetic field contours of opposite polarity. Here we summarize the main aspects of the 3D MHD model for convenience. We use the following normalization of the 3D MHD equations: the coordinates $x\rightarrow x/L_0$, the time $t\rightarrow t/\tau_A$, the velocities $v\rightarrow
v/V_A$, the magnetic field $B\rightarrow B/B_0$, the density $\rho\rightarrow \rho/\rho_0$, and the pressure 
$p\rightarrow p/p_0$, where $L_0$ is the length scale defined below, $V_A$ is the normalizing Alfv\'{e}n speed, $\tau_A
= L_0/V_A$ is the Alfv\'{e}n  time, $B_0$ is the normalizing magnetic field magnitude, $\rho_0$ is the normalizing density,
and $p_0$ is the normalizing pressure in the corona. Other physical parameters are the Lundquist number $S=L_0V_A/\eta$, where $\eta$ is the resistivity  (in the present study we set  
$S=10^5$ and the resistivity has negligible effect on the slow magnetosonic waves), the Froude number  $F_r=V_A^2 L_0/(GM_\odot)$, where $G$ is the gravitational constant and $M_\odot$ is the solar mass, and the Euler number $E_u=p_0/(\rho_0
V_A^2)=C_s^2/\gamma V_A^2$, where $C_s$ is the sound speed. 

\begin{figure}[ht]
\centerline{
\includegraphics[width=\linewidth]{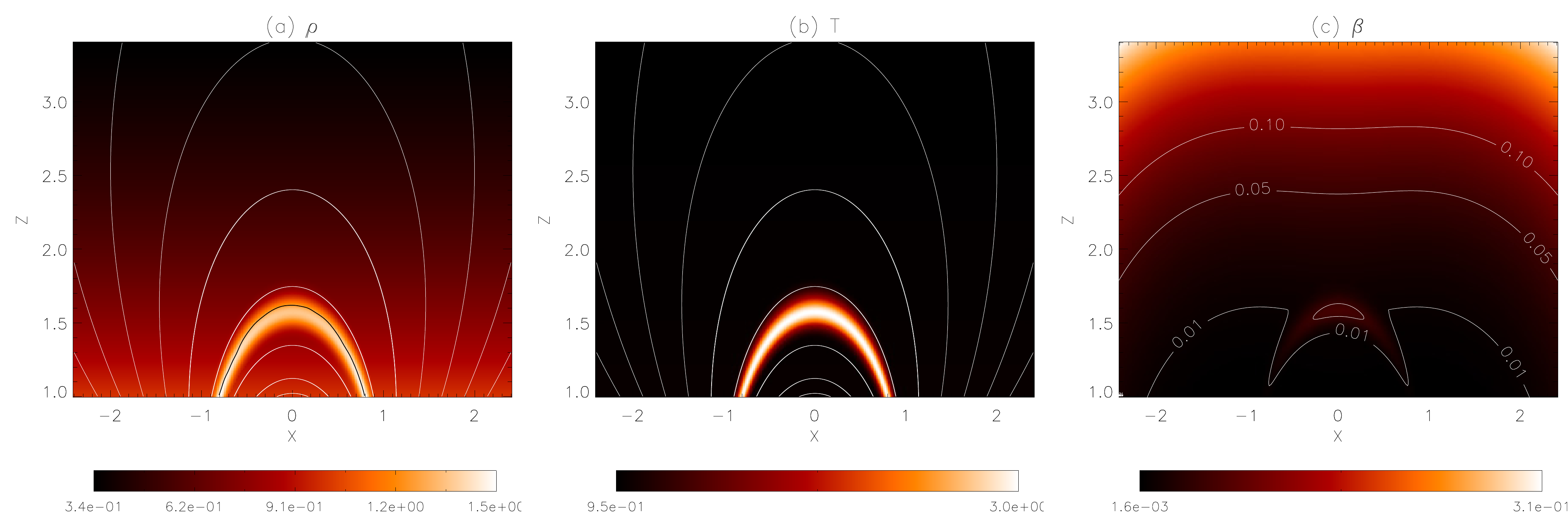}}
\caption{The initial state of the hot and dense coronal loop in the 3D MHD model AR in normalized units shown in the $x-z$ plane cut through the center of the loop. (a) The initial loop density with contrast $n_r=1.5$. Overlaid are the magnetic field lines (white) and a cut along the loop (black). (b) The corresponding initial loop temperature structure with loop-to-corona temperature ratio $T_r=3$ with $T_0=2$ MK. (c) The plasma $\beta$ in the computational domain. The contours show the values of $\beta$=0.01, 0.05, 0.1.} %The contour shows $\beta=1$ location.}
\label{init:fig}
\end{figure}

The initial magnetic field in the present model is a potential dipole \citep[see, e.g.][for the expressions of the Cartesian components of B]{OT02}, and the density is initialized with gravitationally stratified density in polytropic equilibrium with the corresponding coronal temperature, and base density. Guided by observations \citep{Wan15},  the hot loop is initialized along a selected arched flux-tube with footpoints at $z=1$ centered at $x_0=\pm0.8$, $y_0=0$ of a radius $r_0=0.12$. The loop model is specified with the larger density and temperature than in the surrounding corona by a density ratio $n_r=n_{\rm in}/n_{\rm ex}$ and temperature ratio $T_r=T_{\rm in}/T_{\rm ex}$ (see, Table~\ref{param:tab}, where the subscript `in' refers to the peak values in the interior of the loop and `ex' refers to the external adjacent to the loop corona). The temperature and density within the loop's cross section at the footpoints are initialized with Gaussian profiles given by
\begin{eqnarray}
&&\rho_{0r}(r)=(n_r-1)\left(e^{-(2r/r_0)^2}-c_0\right)/(1-c_0)+1,\\
&&T_{0r}(r)=(T_r-1)\left(e^{-(2r/r_0)^2}-c_0\right)/(1-c_0)+1,
\end{eqnarray} where $r$ is the radial distance to the loop's axis, $r_0$ is the loop radius, and the constant $c_0=e^{-4}$ was chosen to match continuously the loop's boundary to the background corona. Taking $\rho_0(x,y,z,r)$ and $T_0(x,y,z,r)$ as the density and temperature at a point ($x,y,z$), passing through which a field line has the footpoint at radial distance $r$ from the axis of the loop, the dependence of the initial background density and temperature on height ($z$) is determined by the gravitational equilibrium and the corresponding polytropic atmosphere given by 
\begin{eqnarray}
&&\rho_0(x,y,z,r)=\left[\rho_{0r}^{\gamma-1}+\frac{H}{c_g c_r}\frac{\gamma-1}{\gamma}\left(\frac1{1+c_g^{-1}(z-1)}-1\right)\right]^{\frac1{\gamma-1}}\\
&& T_0(x,y,z,r)=c_r\rho_0(x,y,z,r)^{\gamma-1},
\end{eqnarray} where $c_r=T_{0r}/\rho_{0r}^{\gamma-1}$, $H=GM_\odot m_p/(2k_BT_0L_0)$ is the normalized  gravitational inverse scale height, where $k_B$ is the Boltzmann constant, and $c_g=10$ is the gravitational length scale ratio parameter where $L_0=R_\odot/c_g$ and $R_\odot$ is the solar radius. This setting of the initial density and temperature distributions implies that the density and temperature along the same field line within the loop have the same contrast to the background corona. With the above setting the loop's footpoint diameter is 10 Mm (measured as the FWHM of the cross sectional density profile) at the coronal base, and the loop length is  $L\sim$149 Mm, measured as the mean length of the extrapolated field lines on the loop boundary (see, Figure~\ref{init:fig}).  In the present study we use an empirical value of the polytropic $\gamma=1.05$ appropriate for the solar corona. Since in hot coronal loops it was found that the polytropic index is close to adiabatic due to the suppression of thermal conduction \citep{Wan15}, we have also preformed test runs with adiabatic $\gamma=5/3$ and found qualitatively similar results, consistent with the fact that the main effect of the viscosity on the wave dissipation is in the momentum equation, not sensitive to the value of $\gamma$. It should be noted that the above initial state is not an exact stable equilibrium, as the transverse pressure gradient in the loop is not balanced by magnetic pressure in the initial state.  However, in the low-$\beta$ plasma (where $\beta$ is the ratio of the thermal to magnetic pressure) the initial state of  the loop disrupts very gradually on a timescale of several hundred Alfv\'{e}n times (evidently, this time scale shortens with increased $T_r$ and $n_r$). We have tested the stability of the initial state by running the model with $T_r=3$ and $n_r=1.5$ without wave injection and found that the loop structure is well preserved to $\sim300\tau_A$ (see the parameters for Case~0 in Table~\ref{param:tab}).

The equations are solved using the well-established modified Lax-Wendroff integration method with 4th order stabilization terms on a uniform grid. The parallelized code is run typical with the resolution of $258^3$ with higher resolution test runs ($258^2\times514$) showing similar results. The code is executed on 256 processors in parallel allowing adequate spatial resolutions with sufficiently small time steps that satisfy the Courant–Friedrichs–Lewy (CFL) condition in the explicit solution method. The CFL condition with viscosity terms taken into account becomes \citep[see, e.g.,][]{SN92}
\begin{eqnarray}
&& \Delta t \leq min\left(c_1\frac{\Delta x}{|V_{p,max}|},c_2\frac{(\Delta x)^2}{\eta_0}\right),
\end{eqnarray} where $V_{p,max}=|V_{max}|+C_{max}$ is the fastest speed of propagation of disturbances, $V_{max}$ is the fastest flow speed, and $C_{max}$ is the fastest phase speed of all wave modes in the computational domain, where $c_1$, and $c_2$ are numerical model-dependent constants of order unity. As one can see, the inclusion of compressive viscosity, may reduce the time step when the inequality  $\frac1{|V_{max}|}>\frac{c_2\Delta x}{c_1\eta_0}$ is satisfied, resulting in considerable limitation on the time step and computational resources for enhanced hot-loop viscosity. Nevertheless, the advantage of the explicit method is the simplicity and high computational efficiency of the parallelized solution of the present visco-resistive 3D MHD model compared to more elaborate implicit methods.

The boundary conditions are open on all external planes of the 3D computational domain, with line-tied boundary conditions at the coronal plane ($z=1$). The slow magnetosonic waves are produced impulsively by a time dependent boundary condition of a velocity pulse injected at the right footpoint of the model coronal loop (centered at $x_0=0.8$, $y_0=0$) with the method similar to past studies \citep{POW18} (see their Equation~(9)-(11)) with the direction of the injected velocity along the magnetic field. The velocity pulse is introduced at the time interval between  $t_1=12$ and $t_2=42$ with the magnitude $V_{0,i}$ given in Table~\ref{param:tab} in each case. The flow injection was modeled along the magnetic field direction $\mbox{\bf B}/|B|$ in the coronal loop as 
\begin{eqnarray}
&&V_i(t) =0.5V_{i,0}\left\{1-cos[2\pi(t-t_1)/t_0]\right\} e^{-\{[(x-x_0)^2+(y-y_0)^2]/r_{i,0}^2\}^2},  t_1<t<t_2,
\end{eqnarray}
where $t_0=30$, and $r_{i,0}=0.7r_0$, with the same form for the associated temperature pulse and normalized amplitude $T_{i,0}=(\gamma-1)V_{i,0}/C_s$.
The time interval of the pulse is chosen to be much shorter than the expected slow magnetosonic wave period in the loop, guided by observations \citep{Wan05,Wan18}.

%4

\begin{table}[h]
\caption{Parameters of the numerical model runs for the various cases in the present study. The normalized injection velocity magnitude $V_{i,0}$, the coronal temperature, and the ratio of the loop temperature $T_r$, and the ratio of the loop density $n_r$ with respect to coronal values. The corresponding normalized coronal values of the compressive viscosity coefficient $\eta_0$: Cases 2 and 5 use the classical values; Cases 3 and 6 use ten times enhanced compressive viscosity coefficients.}
%\begin{center}
\centering
%\vspace{-0.1in}
%\begin{tabular}{ | r | r | r | r |}
\hspace{1cm}\begin{tabular}{cccccc}
\hline\hline
Case \# & $V_{i,0}$ [$V_A$] & $T_0$ [MK]  & $T_r$  & $n_r$ &  $\eta_0$   \\ \hline
	0  &  0.0            &  2       &  3 &  1.5 &  0.0 \\
	1  &   0.0756           &  2       &  3 &  1.5 &  0.0 \\	
    2 &    0.0756          &  2       & 3      &     1.5    &       $7.6\times10^{-5}$    \\
    3 &   0.0756           &  2       & 3           &  1.5    &     $7.6\times10^{-4}$  \\
    4 &    0.1              & 7        & 1.5          &    1.5     &      0.0              \\
    5 &     0.1              & 7        & 1.5          &    1.5     &       $1.6\times10^{-3}$           \\
    6 &     0.1              & 7        & 1.5          &    1.5     &     $1.6\times10^{-2}$      \\
    7 &     0.01              & 7        & 1.5          &    1.5     &       0.0 \\
    \hline
\end{tabular}
\label{param:tab}
\end{table}

%We use the following physical parameters four our model normalization: the magnetic field, $B_0=100$ G, density $n_0=10^9$ cm$^{-3}$, length scale, $L=7\times10^9$ cm, temperature $T_0=7\times10^6$ K, and  the polytropic index $\gamma=1.05$. With these parameters the sound speed is $C_s=348$ km s$^{-1}$, the Alfv\'{e}n speed, $V_A=6897$ km s$^{-1}$, the Alfv\'{e}n time, $\tau_A=10.1$ s, the ratio of the thermal to magnetic pressure, $\beta=0.00486$, and the classical \citet{Bra65} compressive viscosity coefficient $\eta_0=12.9$ g cm$^{-1}$ s$^{-1}$. Note, that the classical compressive viscosity depends primarily on temperature as $\eta_0(T)=\eta_0T^{5/2}$. 
We provide the model parameters of the various cases in the present study in Table~\ref{param:tab}. We use the following physical parameters for our model normalization (see Table~\ref{param:tab}): the magnetic field, $B_0=100$ G, density $n_0=10^9$ cm$^{-3}$, length scale, $L_0=7\times10^9$ cm. With these parameters and temperature $T_0$=2 MK in Cases 0-3, the sound speed is $C_s=182$ km s$^{-1}$, the Alfv\'{e}n speed, $V_A=6897$ km s$^{-1}$, the Alfv\'{e}n time, $\tau_A=10.1$ s, and the classical \citet{Bra65} compressive viscosity coefficient $\eta_0=0.615$ g cm$^{-1}$ s$^{-1}$ in the corona outside the loop. The normalized viscosity coefficient is obtained from $\eta_0\rightarrow \eta_0/(v_AL_0\rho_0)$. In Cases~4-7 we use $T_0$=7 MK, $T_r=1.5$ with corresponding higher ${\mbox \bf \eta_0}$, and the sound speed maximal value 416 km s$^{-1}$ inside the loop. Thus, to keep the injection velocity and waves at similar nonlinearity inside the loop as in Cases~0-3, we set  $V_{i,0}=0.1V_A$ in Cases~4-7. Hence, we use the same ratios of the injected velocities and the peak slow magnetosonic speeds (i.e., the sonic Mach number) for the two modeled temperatures inside the loops in Cases~1-6. Note, that the classical compressive viscosity depends primarily on temperature as $\eta_0(T)\approx\eta_{00}T^{5/2}$, with some weak dependence of the Coulomb logarithm on $T$. Thus, for a loop with $T_r=1.5$ the compressive viscosity at the loop axis peaks with $\sim2.8$ times the coronal value, and for $T_r=3$ the compressive viscosity peaks with $\sim16$ times the coronal value. In the present study we compare inviscid modeling results with the classical values of $\eta_0$ modeling results, and ten-times enhanced viscosity cases. 
%
%\begin{eqnarray}
%V_z(x,z=0,t)=\frac{V_0}{2}(cos\omega t+1)e^{-[(x-x0)/2w]^2},
%\end{eqnarray} where the amplitude of the velocity perturbation $V_0$ is few percent of the coronal Alfv\'{e}n speed, and the frequency of the waves, $\omega$ correspond to periods of several minutes, and $x_0=0$ is at the center of the prominence foot. The particular form of the velocity perturbations provides upward propagating waves.

\newpage\section{Numerical Results} \label{num:sec}

\begin{figure}[ht]
\centerline{
\includegraphics[width=0.8\linewidth]{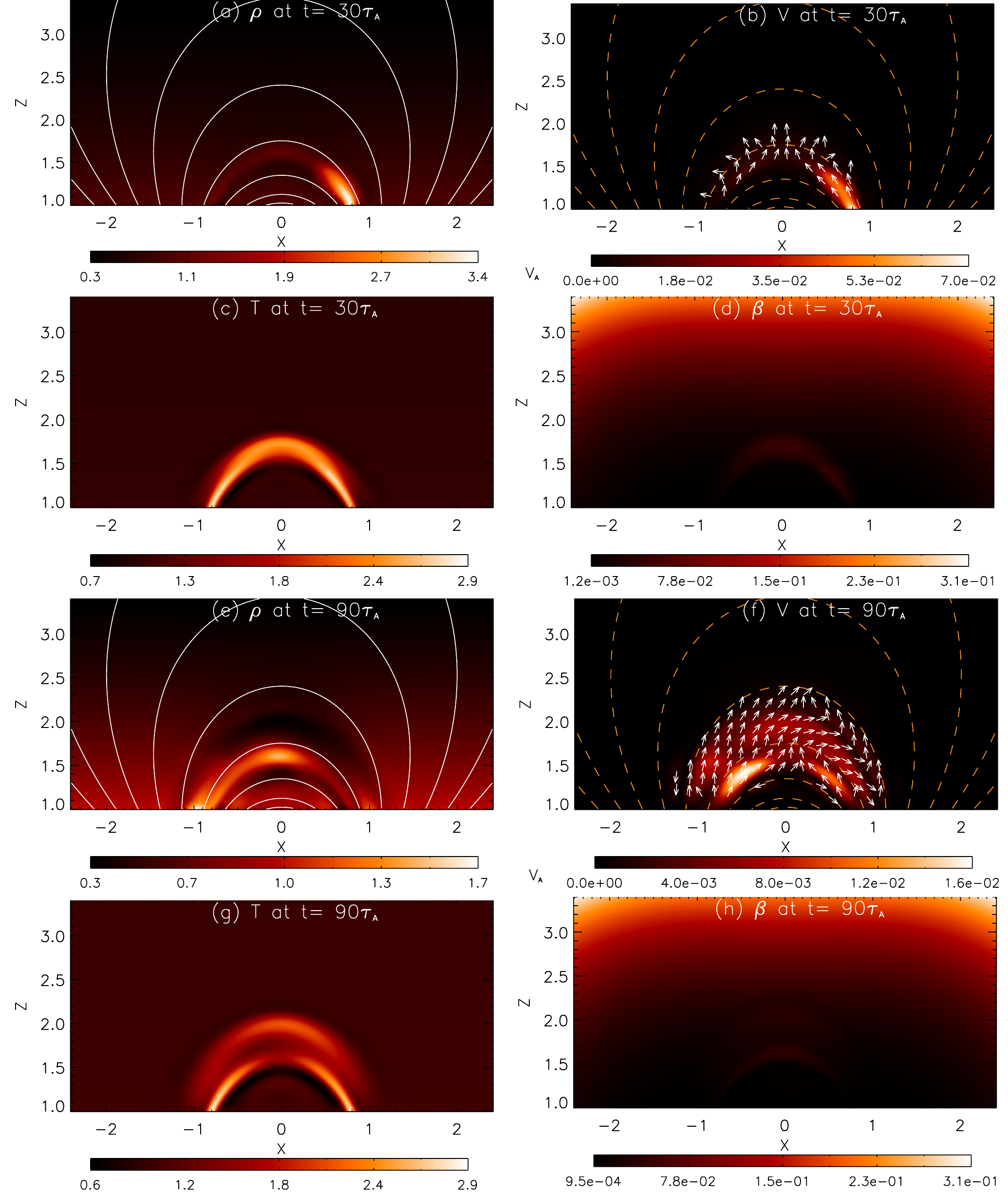}
}
\vspace{-0.25cm}\caption{The results of the 3D MHD model of the slow magnetosonic wave injection into the coronal loops in the initial and later stages of the evolution for Case~2 with $T=2$ MK, $T_r=3$, $n_r=1.5$, and $\eta_0=7.6\times10^{-5}$. The snapshot of the variables in the $x-z$ plane at $y=0$ at two times of the evolution. (a) The density $\rho$ at $t=30\tau_A$, the white lines show several magnetic field lines. (b) The velocity magnitude $v$ with arrows indicating the direction of the velocity vectors for $|v|>0.06v_{max}$, where $v_{max}$ is the maximal velocity in the plane. (c) The temperature $T$, (d) the plasma $\beta$. (e)-(h): Same as (a)-(d) but for the cases at time $t=90\tau_A$. The animation of $\rho$ (panels a, e) and $v$ (panels b, f) for times 0-312$\tau_A$ is available online.}
\label{vart_case2:fig}
\end{figure}

\begin{figure}[ht]
\centerline{
\includegraphics[width=0.7\linewidth]{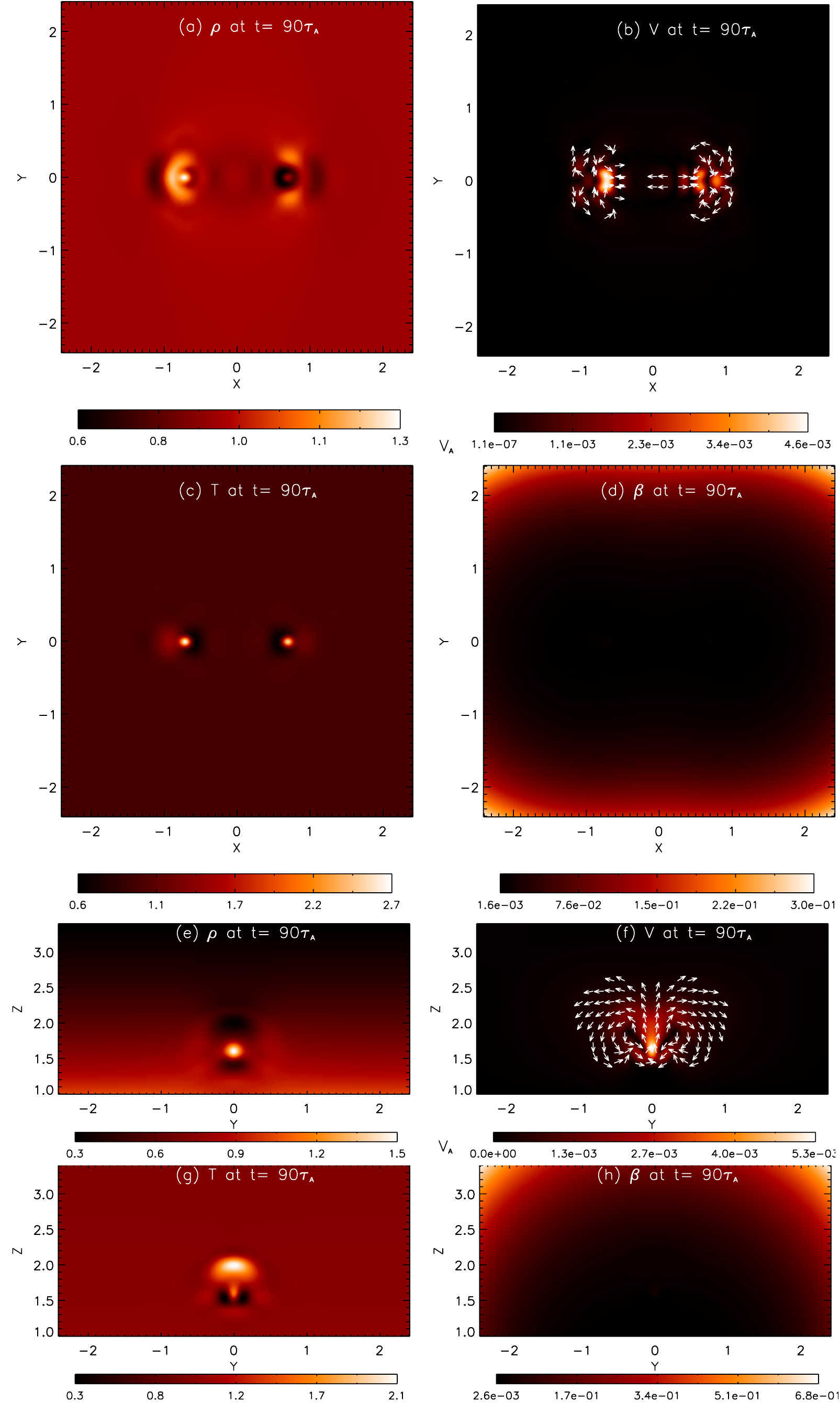}
}
\caption{The results of the 3D MHD model of the slow magnetosonic wave injection into the coronal loop for Case~2 with $T=2$ MK, $T_r=3$, $n_r=1.5$, and $\eta_0=7.6\times10^{-5}$ at $t=90\tau_A$. The snapshot of the variables in the $x-y$ plane at $z=1.18$: (a) the density $\rho$, (b) the velocity magnitude $v$ with arrows indicating the direction of the velocity vectors, (c) the temperature $T$, (d) the plasma $\beta$. The variables in the $y-z$ plane at $x=0$:  (e) $\rho$ (f) $v$  (g) $T$,  (h) $\beta$. The animations of $\rho$ (panels a, e) and $v$ (panels b, f) for times 0-312$\tau_A$ are available online.}
\label{vart_yz_xy_case2:fig}
\end{figure}

\begin{figure}[ht]
\centerline{
\includegraphics[width=\linewidth]{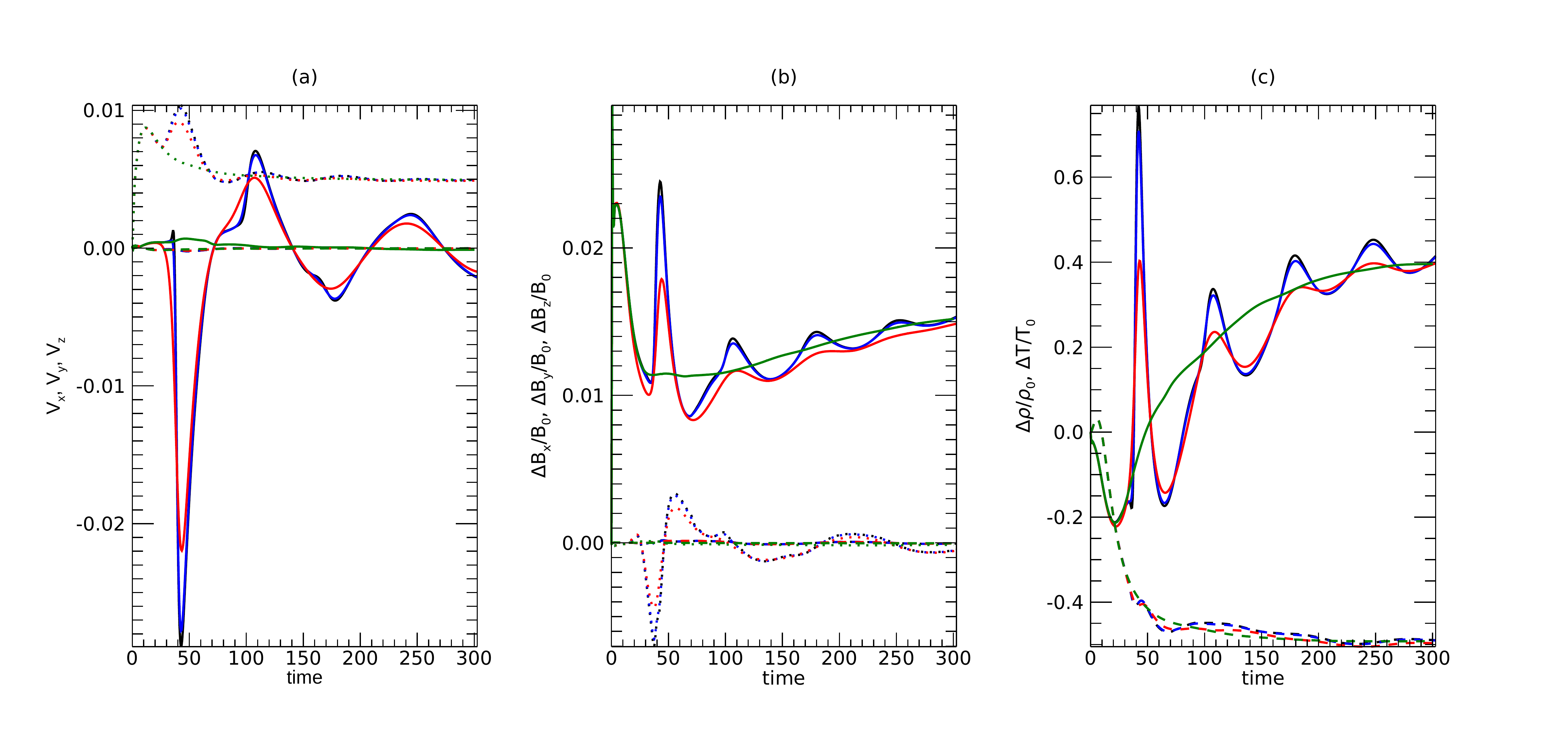}
}
\caption{The temporal evolution of the variables near loop apex at $x=0$, $y=0$, $z=1.6$ for Cases~1 (black), Case~2 (blue) and Case~3 (red). Note, that Case 1 and 2 are nearly identical. Case~0 (green) without flow injection is shown for reference. (a) The velocity components $V_x$ (solid), $V_y$ (long dashes), and $V_z$ (short dashes). (b)  The normalized magnetic field component perturbations $\Delta B_x/B_0$ (solid), $\Delta B_y/B_0$ (long dashes), and $\Delta B_z/B_0$ (short dashes), where $B_0$ is the magnetic field magnitude at $t=0$ at the plotted location. (c) The normalized density perturbation $\Delta\rho/\rho_0$ (solid) and temperature perturbation $\Delta T/T_0$ (long dashes) where $\rho_0$ and $T_0$ are the values at $t=0$ at the plotted location.}
\label{vdbdnT_Tr3:fig}
\end{figure}

\begin{figure}[ht]
\centerline{
\includegraphics[width=0.95\linewidth]{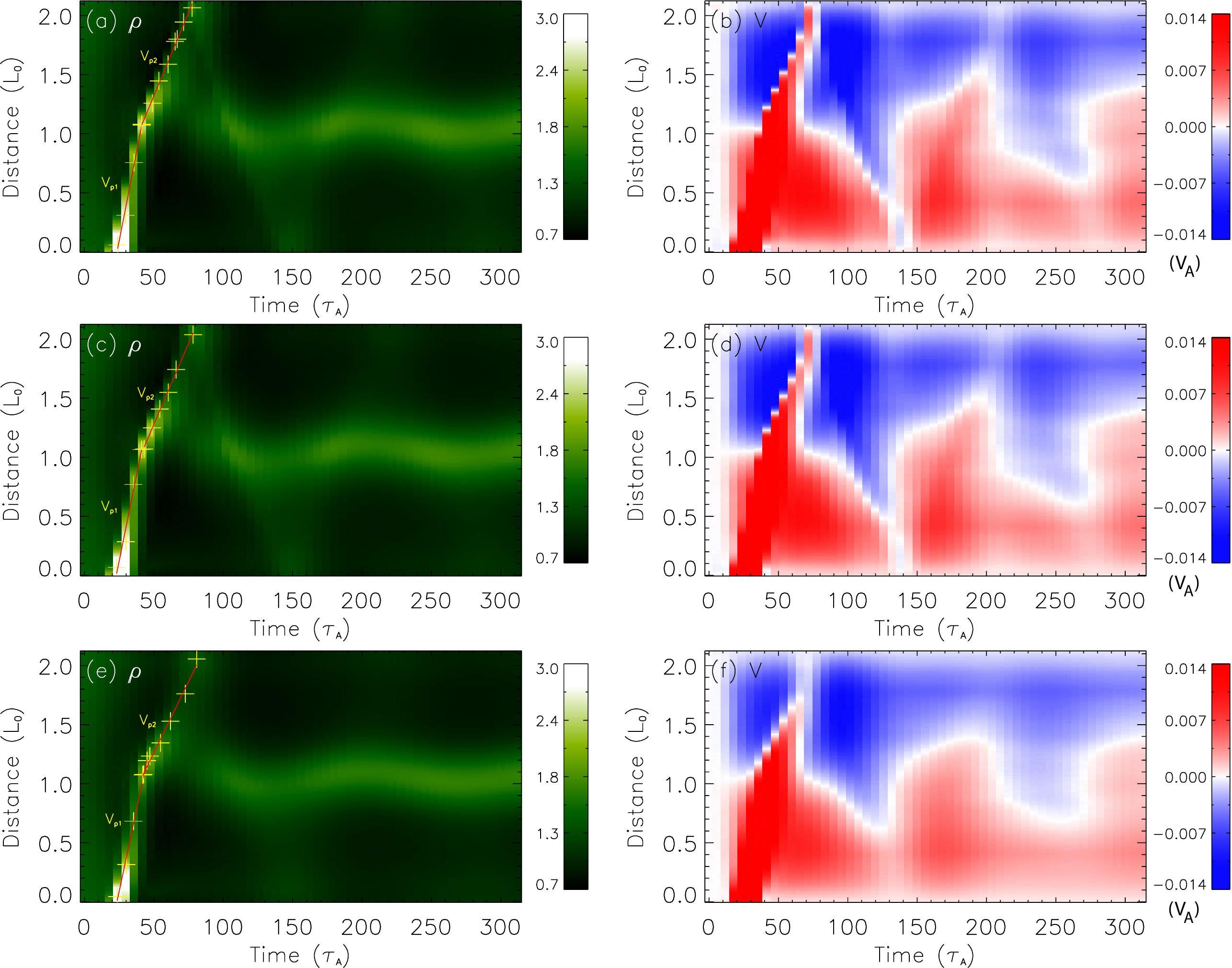}
}
\caption{Space-time plots of the density (left panels) and the velocity component parallel to the magnetic field (right panels) for a cut along the loop as shown by the black line in Figure~\ref{init:fig}a due to injection of a flow pulse. (a) $\rho$ and (b) $V$ for Case 1 without viscosity. (c) and (d): Same as (a) and (b) but for Case 2 with $\eta_0=7.6\times10^{-5}$. (e) and (f): Same as (a) and (b) but for Case 3 with $\eta_0=7.6\times10^{-4}$. In left panels, we measure the wave propagation speeds by linear fits (red lines) to the marked points (pluses) on the bright ridges (see text in detail).  In left panels, the yellow crosses and the red lines were  used to calculate propagation velocity of the perturbations.}
%The space-time plot of the density (top panels) and velocity (lower panels) along the loop due to the flow injection for (a) Case 1 without viscosity, (b) Case 2 with $\eta_0=3.8\times10^{-5}$, (c) Case 3 with $\eta_0=3.8\times10^{-4}$. }
\label{xt_Tr3:fig}
\end{figure}

\begin{figure}[ht]
\centerline{
\includegraphics[width=0.8\linewidth]{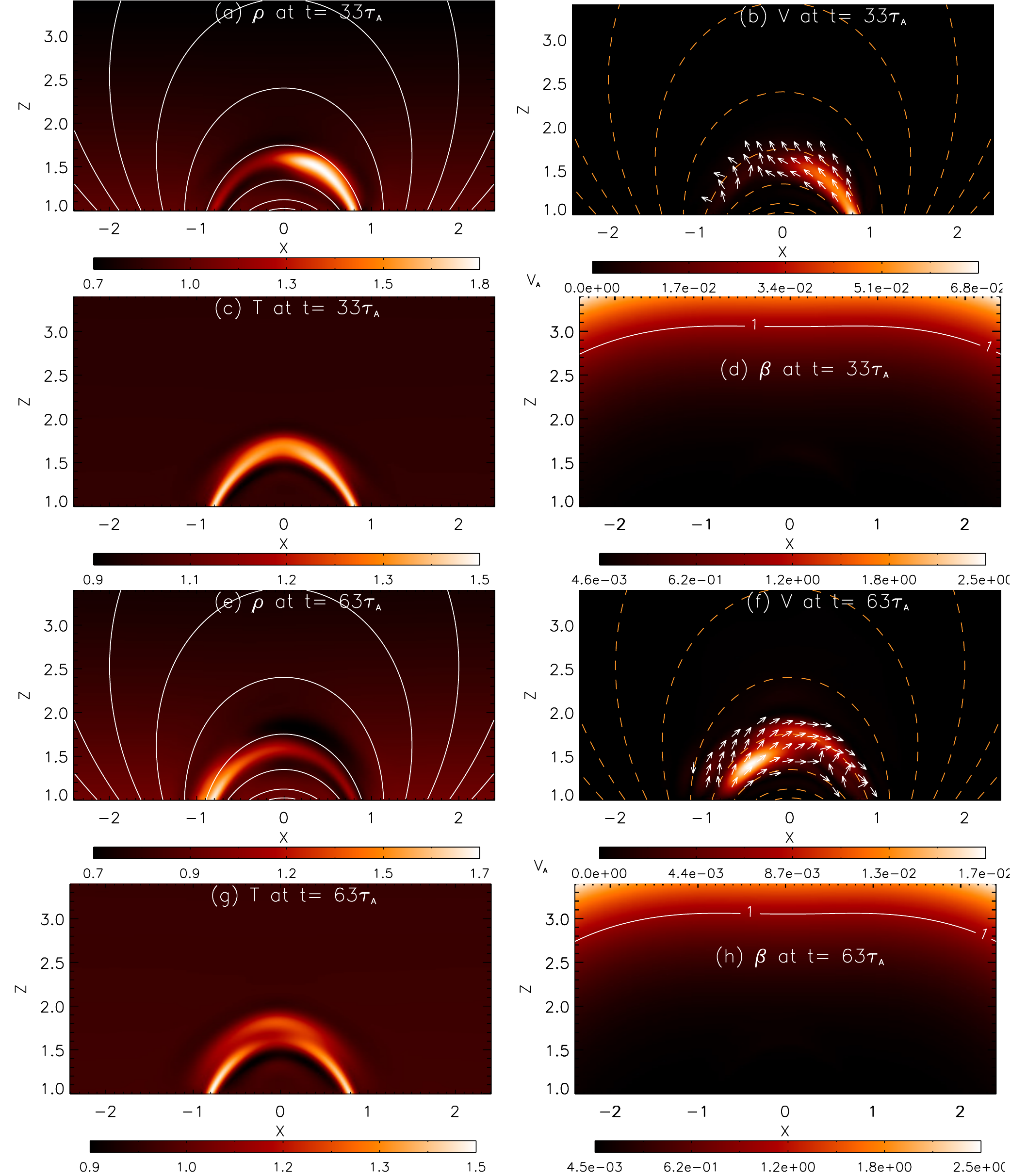}}
\caption{The results of the 3D MHD model of the slow magnetosonic wave injection into the coronal loops in the initial and later stages of the evolution for Case~5 with $T=7$ MK, $T_r=1.5$, $n_r=1.5$, and $\eta_0=1.6\times10^{-3}$. The snapshot of the variables in the $x-z$ plane at $y=0$ at two times of the evolution. (a) The density $\rho$ at $t=33\tau_A$, the white lines show several magnetic field lines. (b) The velocity magnitude $v$ with arrows indicating the direction of the velocity vectors for $|v|>0.06v_{max}$, where $v_{max}$ is the maximal velocity in the plane. (c) The temperature $T$, (d) the plasma $\beta$. At time $t=63\tau_A$ (e) same as (a), (f) same as (b), (g) same as (c), (h) same as (d). The animation of of $\rho$ (panels a, e) and $v$ (panels b, f) for times 0-312$\tau_A$ are available online.}
\label{vart_case5:fig}
\end{figure}

\begin{figure}[ht]
\centerline{
\includegraphics[width=0.65\linewidth]{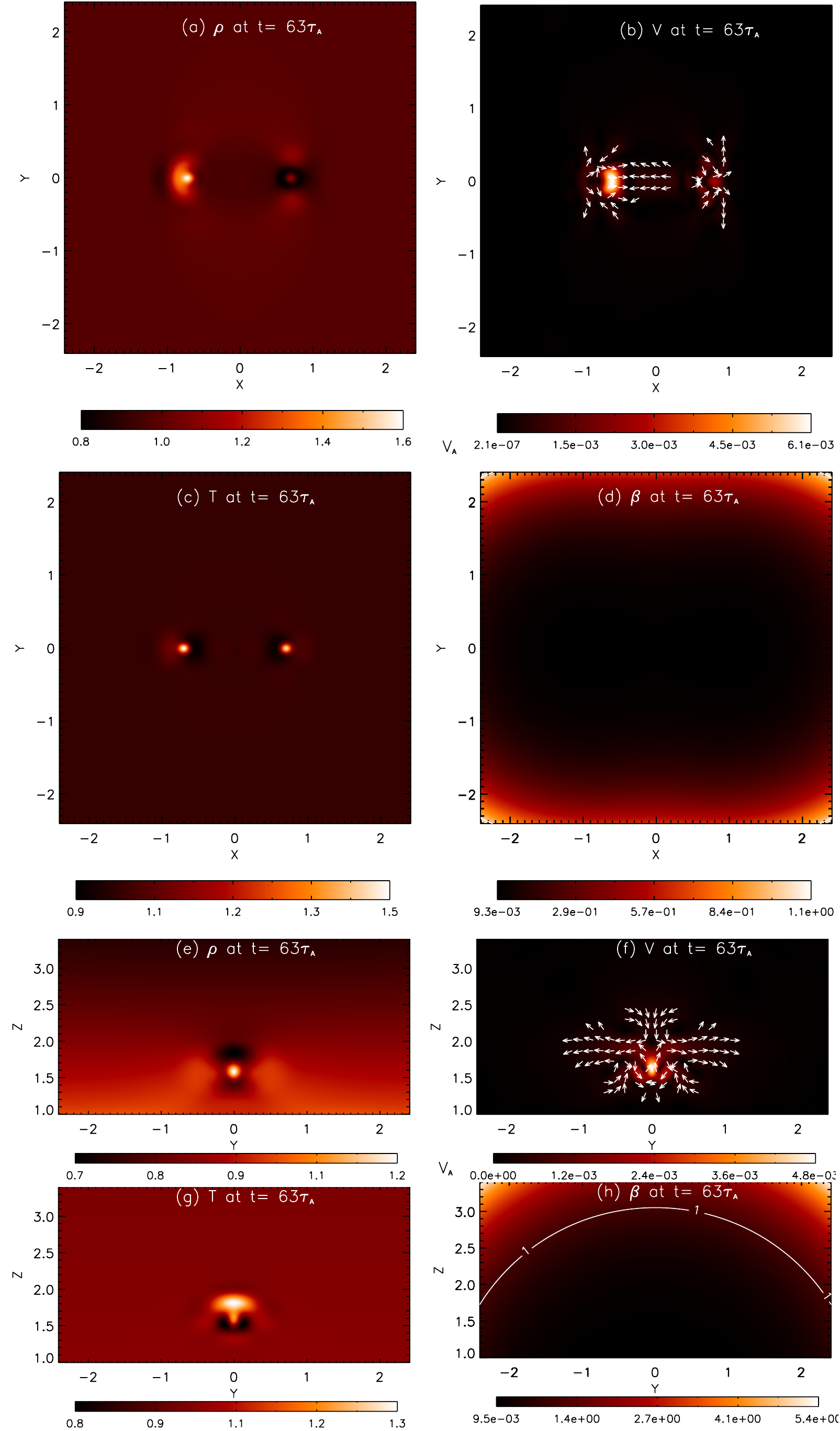}}
\caption{The results of the 3D MHD model of the slow magnetosonic wave injection into the coronal loops for Case~5 with $T=7$ MK, $T_r=1.5$, $n_r=1.5$, and $\eta_0=1.6\times10^{-3}$ at $t=63\tau_A$. The snapshot of the variables in the $x-y$ plane at $z=1.18$: (a) the density $\rho$, with the white lines showing several magnetic field lines. (b) The velocity magnitude $v$ with arrows indicating the direction of the velocity vectors. (c) The temperature $T$. (d) The plasma $\beta$. The variables in the $y-z$ plane at x=0: (e) $\rho$, (f) $v$,  (g) $T$,  (h) $\beta$. The animations of $\rho$ (panels a, e) and $v$ (panels b, f) for times 0-312$\tau_A$ are available online.}
\label{vart_yz_xy_case5:fig}
\end{figure}

\begin{figure}[ht]
\centerline{
\includegraphics[width=\linewidth]{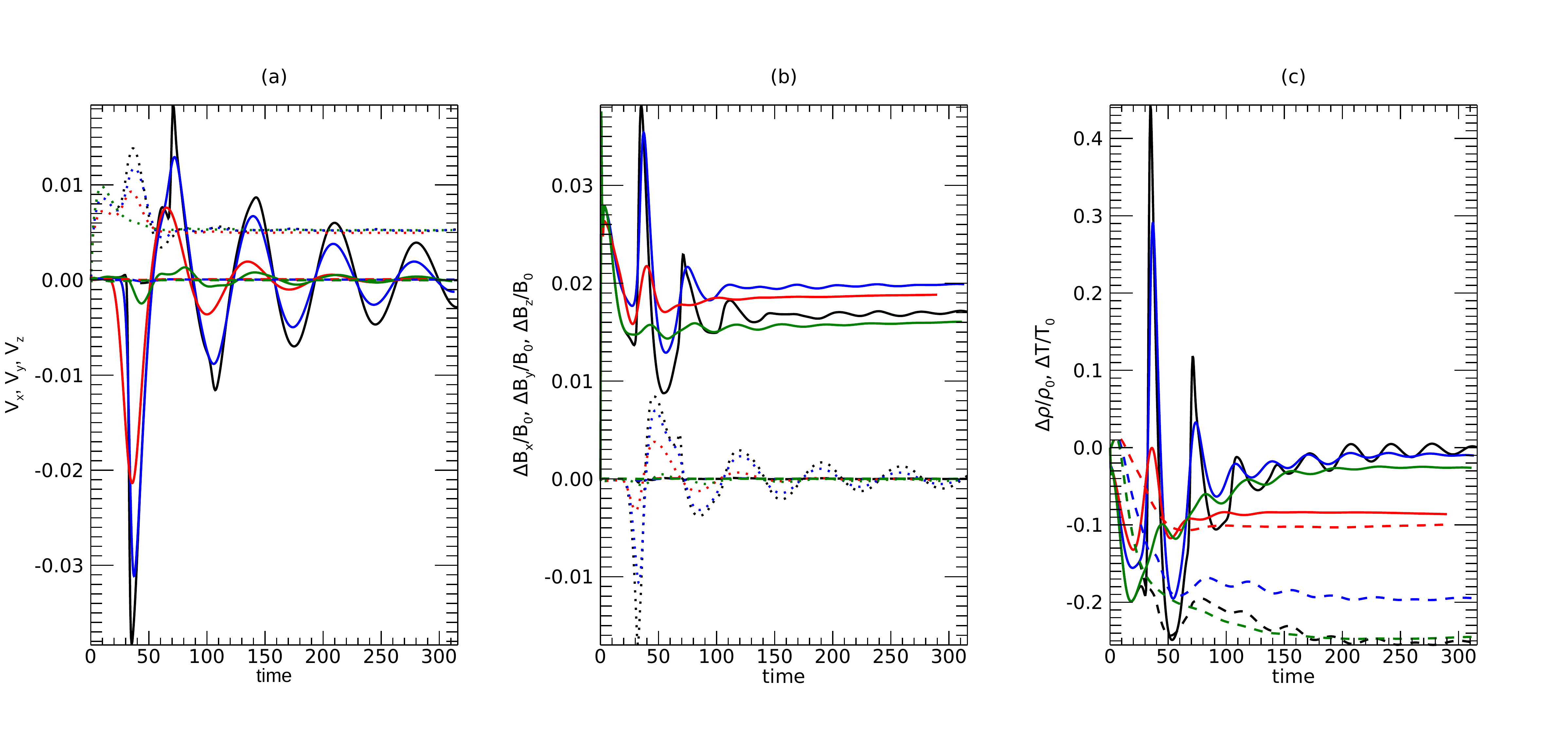}
}
\caption{The temporal evolution of the variables near loop apex at $x=0$, $y=0$, $z=1.6$ for Cases 4 (black), Case 5 (blue), Case 6 (red), Case 7 (green). (a) The velocity components $V_x$ (solid), $V_y$ (long dashes), and $V_z$ (short dashes). (b)  The normalized magnetic field component perturbations $\Delta B_x/B_0$ (solid), $\Delta B_y/B_0$ (long dashes), and $\Delta B_z/B_0$ (short dashes), where $B_0$ is the magnetic field magnitude at the plotted location. (c) The normalized density perturbation $\Delta\rho$ (solid) and temperature perturbation $\Delta T$ (long dashes) where $\rho_0$ and $T_0$ are the values at $t=0$ at the plotted location.}
\label{vdbdnT_T7Mk:fig}
\end{figure}

\begin{figure}[ht]
\centerline{
\includegraphics[width=\linewidth]{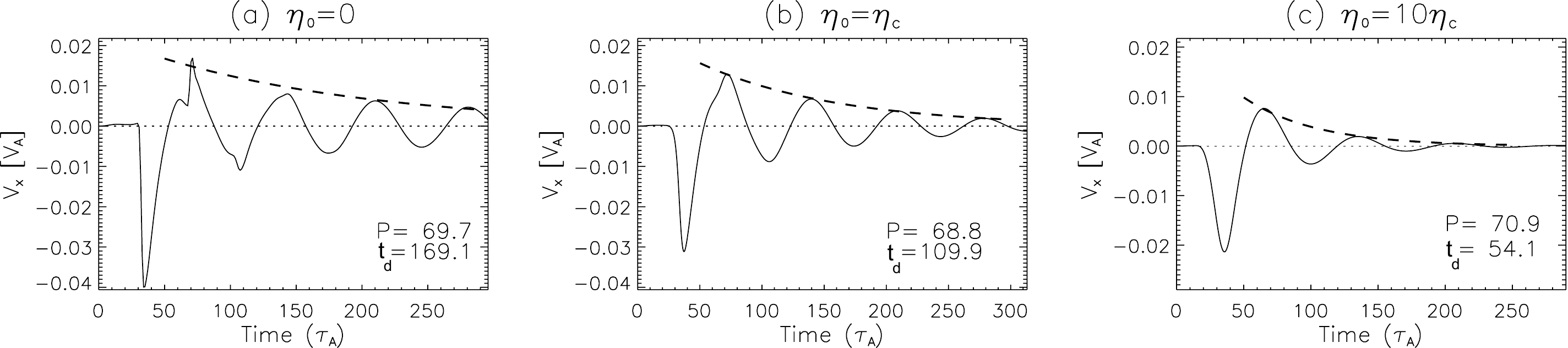}
}
\caption{Measurements of the oscillation parameters for $V_x$ near loop apex by fitting the envelope with an exponential decay function for the cases with $\eta_0=0$ (a), $\eta_0=\eta_c$ (b), and $\eta_0=10\eta_c$ (c), where $\eta_c$ represents the classical viscosity coefficient. The dashed line indicates the best fit exponential decay to the peaks of $V_x$.  The values of the period, $P$, and the damping times, $t_d$, are indicated on the panels.}
\label{fig:vxfit}
\end{figure}

\begin{figure}[ht]
\centerline{
\includegraphics[width=0.95\linewidth]{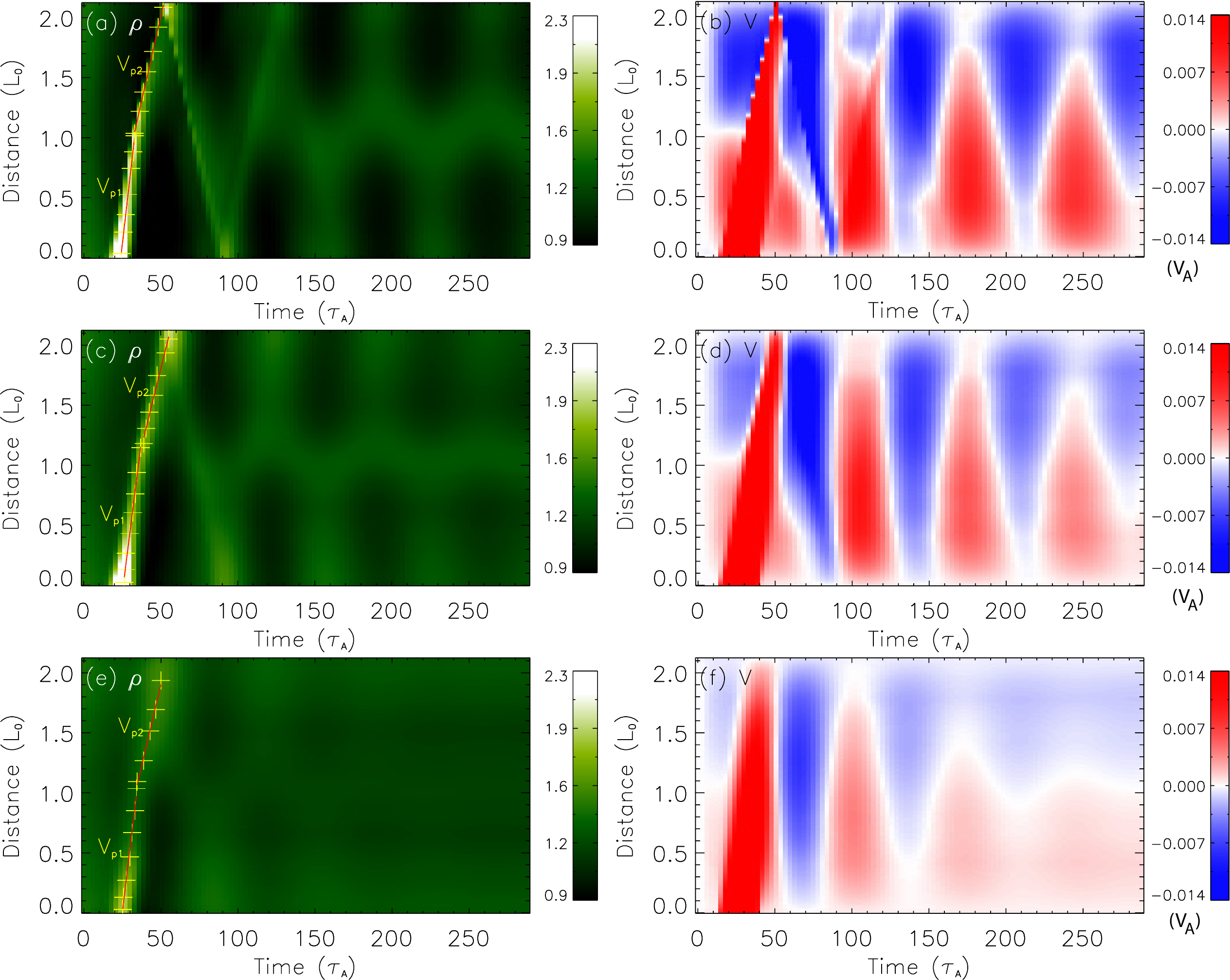}
}
\caption{Space-time plots of the density (left panels) and the velocity component parallel to the magnetic field (right panels) for a cut along the loop as shown in Figure~\ref{init:fig}a (marked with black line) due to injection of a flow pulse. (a) $\rho$ and (b) $V$ for Case 4 without viscosity. (c) and (d): Same as (a) and (b) but for Case 5 with $\eta_0=1.6\times10^{-3}$. (e) and (f): Same as (a) and (b) but for Case 6 with $\eta_0=1.6\times10^{-2}$. In left panels, we measure the wave propagation speeds by linear fits (red lines) to the marked points (pluses) on the bright ridges (see text in detail). In left panels, the yellow crosses and the red lines were  used to calculate propagation velocity of the perturbations.}
\label{xt_T7Mk:fig}
\end{figure}

In Figures~\ref{vart_case2:fig}-\ref{xt_T7Mk:fig} we present the results of the 3D MHD modeling of waves excited in a hot and dense coronal loop by a velocity pulse at the footpoint for Cases~1-7 with the parameters summarized in Table~\ref{param:tab}. 

%5
In Figure~\ref{vart_case2:fig} we show the the snapshot of the variables in the $x-z$ plane at $y=0$ (the plane of the loop axis) at two times for Case~2 with the parameters $T=2$ MK, $T_r=3$, $n_r=1.5$, and $\eta_0=7.6\times10^{-5}$. The coronal loop density $\rho$ is shown in Figure~\ref{vart_case2:fig}a at $t=30\tau_A$ overlayed with several representative magnetic field lines. The enhanced density due to the velocity pulse injection is evident above the right footpoint of the loop. The corresponding velocity magnitude $v$ due to the pulse is presented in Figure~\ref{vart_case2:fig}b, where the fixed-length arrows indicated the local direction of the velocity vectors. The enhanced loop temperature $T$, and the plasma $\beta$ are shown in Figure~\ref{vart_case2:fig}c and \ref{vart_case2:fig}d, respectively. The temperature increase due to the pulse is small relative to the background hot-loop temperature mainly because of the choice of $\gamma$=1.05. Thus, the main contribution to $\beta$ increase in the loop is due to the density pulse. The evolution of the pulse and the reflected slow magnetosonic wave is evident in Figures~\ref{vart_case2:fig}e-\ref{vart_case2:fig}g at $t=90\tau_A$ after reflection from the left footpoint of the loop. In addition to the reflection, there is a gradual evolution in the background state due to the nonequilibrium initial state of the loop. The animation of the density in the $x-z$ plane cut at $y=0$ due to the injection of the pulse and the generation of the waves for Case~2 is available online.

%6
Figure~\ref{vart_yz_xy_case2:fig} shows the cut in the planes perpendicular to the plane of the loop with peak temperature $T_{\rm max}$=6 MK inside the loop.   The variables in $x-y$ plane at $z=1.18$ and at the $y-z$ plane at $x=0$ for Case~2 are shown. The $x-y$ plane is located near the footpoints of the loop at $z=1.18$ exhibiting the density, velocity, temperature, and $\beta$ at $t=90\tau_A$. The transverse waves generated near the footpoints by the flow pulse are evident in the density variability outside the loop (see the related animation online) in Figure~\ref{vart_yz_xy_case2:fig}a. The flows associated with the transverse oscillations are shown in Figure~\ref{vart_yz_xy_case2:fig}b, and the temperature variability in Figure~\ref{vart_yz_xy_case2:fig}c. The values of $\beta\ll1$ are small in this plane near the footpoints of the loop. The $y-z$ plane at $x=0$ cuts across the loop apex and shows the oscillations and the associated velocities outside the loop. The larger scale changes in the loop density and temperature, and the associated velocities are due to gradual changes in the background loop structure, associated with small changes in the magnetic structure and the effects of the velocity pulse. The small magnetic field changes produce amplified changes of the thermal pressure in the low-$\beta$ plasma, as evident from the pressure balance condition (in normalized units) $B^2+\beta p=const.$  Thus, the small amplitude fast magnetosonic perturbations carry a significant  energy flux of the slow magnetosonic waves in the low-$\beta$ plasma, evident by considering the linearized perturbation of the pressure balance condition $2\delta B B_0+\beta \delta p=0$. It is evident that $\Delta B_z/B_0$ and $V_x$ of the wave are phase shifted by half period and are related by a factor of two in amplitude after the initial pulse (see, Figures~\ref{vdbdnT_Tr3:fig} and \ref{vdbdnT_T7Mk:fig}). This suggest that the vertical kink mode is driven by the slow mode wave (see the discussion in \citet[][]{Ofm12}). Thus, quantitative estimate of the slow/fast mode coupling is required to improve coronal seismology application to the determination of the dissipation coefficients. This coupling could depend on several factors, such as the details of the loops' structure, the magnitude of the pulse (i.e., nonlinearity), and the temperature of the loop. The present results are in qualitative agreement with previous finding of \citet{Ofm12} that demonstrated the mode couplings and the generation of both, slow and fast mode oscillations by velocity pulse in hot coronal loops in more simplified AR model. The main differences are due to the crossectional loop structure, the details of the wave excitation, and compressive viscosity in the present study.

%7

We compare the temporal evolution of the variables at the loop apex ($x=0$, $y=0$, $z=1.6$) for the inviscid case (Case~1), with classical compressive viscosity coefficient $\eta_0=7.6\times10^{-5}$ in normalized units (Case~2), and with ten times enhanced compressive viscosity coefficient $\eta_0=7.6\times10^{-4}$ for the $T_{\rm max}$=6 MK loop case in Figure~\ref{vdbdnT_Tr3:fig}. For reference the evolution due to the initial background adjustment and the gradual change without flow injection (Case~0) is shown. Note, that inside the hot loop the values of the compressive viscosity coefficient increases by a factor $T_r^{5/2}=3^{2.5}=15.6$ at the loop axis. It is evident  in Figure~\ref{vdbdnT_Tr3:fig}a that the velocity component along the loop at the apex ($V_x$) shows oscillatory evolution, with initial peak due to the arrival of the pulse at the loop apex, followed by damping of the wave amplitude, with the inviscid case evolution nearly identical to the case with classical viscosity, and with evident increased dissipation for the enhanced viscosity case. The main period of the velocity oscillations is $\sim 150\tau_A$, while the magnetic field perturbations $\Delta B_x$ and density oscillations show about half of the period of $\sim80\tau_A$ at the apex. The wave mode and the frequency doubling found in the magnetic field and density is the nonlinearly driven vertical kink mode oscillation, consistent with the discussion in \citet{Ofm12}. The damping time is on the order of one oscillation period, with dominant damping in the initial nonlinear stage of the oscillations and with similar damping time for the inviscid, classical viscosity, and enhanced viscosity in Cases~1-3. The calculated values of the damping time from an exponential fit to the two peaks of $V_x$  are $\sim131\tau_A$ for $\eta_0=7.6\times10^{-5}$ (nearly identical to the inviscid case), and $\sim127\tau_A$ for  $\eta_0=7.6\times10^{-4}$ with similar value for the inviscid case. Since the flow injection is along the loop, the largest velocity components are in the $x-z$ plane, and due to symmetry of the loop with respect to $y=0$ plane, the $V_y$ component is smaller than the $V_x$ and $V_z$ components everywhere in the loop. The velocity and density oscillations show weak dependence on $\eta_0$ at the apex, while the small magnetic field component perturbations $\Delta B_x$ (Figure~\ref{vdbdnT_Tr3:fig}b) show the clearest (albeit small) dependence on $\eta_0$. This is likely due to the coupling of the slow magnetosonic wave produced by the pulse with nonlinear fast magnetosonic oscillations through variation of total (thermal and magnetic) pressure perturbation inside the loop. The evolution of the density perturbation $\Delta\rho$ and temperature perturbation $\Delta T$ is shown in Figure~\ref{vdbdnT_Tr3:fig}c. We find that the ratio of the magnetic and thermal pressure perturbations (in normalizing units) is consistent with the small value of $\beta\sim10^{-2}$, as expected from the pressure balance. This can be shown by using the normalized pressure balance condition $B^2+\beta p=const.$ and deriving the ratio of the first order perturbations of the magnetic and thermal pressures around the pressure balance. Thus, we get the relations $\Delta{(B^2)}/\Delta{p}=2(\Delta{B_x}+\Delta{B_y}+\Delta{B_z})/(\Delta{\rho}+\Delta{T})\sim\beta$, where we have used $B=1$, $T=1$, and $\rho=1$ for the normalized background quantities. We note that since the plasma is nearly isothermal $\Delta T\approx 0$ can be neglected compared to $\Delta \rho$ in the above expression. In addition to the wave, the temperature is affected by the non-periodic change in the background equilibrium, as evident from the decrease of the temperature at the apex of the loop (see, Figure~\ref{vart_case2:fig}g).

%8
The space-time plots of Case~1 (inviscid), Case~2 with  classical compressive viscosity $\eta_0=7.6\times10^{-5}$ and Case~3 with an order of magnitude enhanced compressive viscosity $\eta_0=7.6\times10^{-4}$ are shown in Figure~\ref{xt_Tr3:fig}, where the density (left panels) and velocity (right panels) perturbations move along the loop for the three cases. The onset of the pulse is at $t=12\tau_A$, and  reflection of the wave is evident at $t\sim75\tau_A$, with the propagation speed is evident from the slope of the linear fits (red lines) to the model data point (yellow crosses).
%close to the sonic speed (in normalized units) for the hot loop with $T_r=3$ (that corresponds to 6 MK in physical units). 
It is evident that the initial disturbances show  distinct propagation speeds. From the linear best fits, we obtain the speeds $V_{p1}=414\pm25$ km~s$^{-1}$ and $V_{p2}=194\pm6$ km~s$^{-1}$ for Case 1, $V_{p1}=401\pm49$ km~s$^{-1}$ and $V_{p2}=189\pm5$ km~s$^{-1}$ for Case 2, and $V_{p1}=383\pm18$ km~s$^{-1}$ and $V_{p2}=166\pm9$ km~s$^{-1}$ for Case 3. We find that the initial upward propagation speeds are larger than the maximum sound speed ($\sim$323 km~s$^{-1}$) at $T_{\rm max}$=6 MK while the downward propagation speeds are close to the background $C_s$ ($\sim$187 km~s$^{-1}$) at $T_0$=2 MK. The slightly supersonic, initial shocked wave front is be formed by injected flow when traveling from the footpoint towards the loop apex{, as evident in Figures~\ref{vart_case2:fig}a, b and the associated animations}. We also notice that in the case with the enhanced viscosity (Case~3) the wave front propagates at slightly slower speed (by about 8\%-17\%) than the case without viscosity (Case~1), since the shock front has dissipated. The damping of the wave is evident from the decrease of the intensity of the density and velocity perturbations, where the dominant effect is leakage and the classical viscosity has small effect, with more pronounced viscous damping in Case~3 with enhanced viscosity. 
%The change in the background loop structure {\bf may} also contributes to the leakage of the propagating wave due to the reduced density and temperature contrasts that reduces wave trapping, as evident in later stages of the evolution, following the first reflection of the wave. 

%9
Figure~\ref{vart_case5:fig} shows the the snapshot of the variables in the $x-z$ plane at $y=0$ through the loop axis at two times for Case~5 of a hotter loop (than in Cases~0-4) with $T=7$ MK, $T_r=1.5$, $n_r=1.5$, and correspondingly higher classical compressive viscosity coefficient $\eta_0=1.6\times10^{-3}$ in normalized units. The coronal loop density $\rho$ is shown in Figure~\ref{vart_case5:fig}a at $t=33\tau_A$ overlayed with several representative magnetic field lines. The enhanced density due to the velocity pulse injection is evident above the right footpoint of the loop as it propagates towards the apex. The larger $C_s$ in the loop in Case~5 compared to Case~2 results in faster propagation of the disturbance (note, that the injection velocity amplitude is similar in both cases in terms of the two $C_s$'s). The corresponding velocity magnitude $v$ due to the pulse is shown in Figure~\ref{vart_case5:fig}b, where the fixed-length arrows indicating the local direction of the velocity vectors. The enhanced loop temperature $T$, and the plasma $\beta$ are shown in Figure~\ref{vart_case5:fig}c and \ref{vart_case5:fig}d, respectively. It is evident that the temperature increase due to the pulse is small relative to the background hot-loop temperature, and the $\beta$ increase in the loop is small, and therefore difficult to distinguish from the background corona in the larger $\beta$ case, that reaches values $\beta>1$ (indicated by the white contour) in the top part of the model AR. The evolution of the pulse and the reflected slow magnetosonic wave is evident at later time $t=63\tau_A$, shown in Figures~\ref{vart_case5:fig}e-\ref{vart_case5:fig}g following reflection from the left footpoint of the loop. Here as well we see  gradual evolution of the background state of the loop as the initial state is not quite in equilibrium. The animation of the density in the $x-z$ plane cut at $y=0$ due to the injection of the pulse and the generation of the waves for Case~5 is available online.

%10
In Figure~\ref{vart_yz_xy_case5:fig} we show the cuts in the planes perpendicular to the plane of the hotter loop axis with peak temperature $T_{\rm max}$=10.5 MK on the loop axis. The $x-y$ plane at $z=1.18$ and the $y-z$ plane at $x=0$ of the variables are shown for Case~5. The $x-y$ plane near the footpoints of the loop at $z=1.18$ shows the density, velocity, temperature, and $\beta$ at $t=63\tau_A$. The transverse compressible fast magnetosonic waves generated near the footpoints by the slow mode pulse are evident in the density perturbations outside the loop (see the related animation online) in Figure~\ref{vart_yz_xy_case5:fig}a. The velocities associated with the transverse fast magnetosonic oscillations are shown in Figure~\ref{vart_yz_xy_case5:fig}b, and the spatial perturbation of the temperature is evident. The values of $\beta\ll1$ are small in this plane near the footpoints of the loop compared to the peak values of $\beta$ in the model AR outside the loop. The $y-z$ plane at $x=0$ that cuts across the loop apex shows the oscillations and the associated velocities outside the loop. The modification of the initial loop density and temperature background structure is evident, and the associated flows are due to the disruption of the near-equilibrium initial state, in combination with the nonlinear effect of the velocity pulse that leads to gradual changes in the loop structure, associated with small changes in the magnetic structure. These results as well are in qualitative agreement with previous finding of \citet{Ofm12} that demonstrated the mode couplings and the generation of both slow and fast mode oscillations by velocity pulse in coronal loops, where the present compressive viscosity damps some of the wave oscillations, while the present model loop cross-sectional structure improves the the wave trapping.

%11

%12

We compare the temporal evolution of the variables at the loop apex ($x=0$, $y=0$, $z=1.6$) for the hotter loop in Cases~4-6 with $T_{\rm max}$=7 MK and $T_r=1.5$ for the inviscid Cases~4, with classical compressive viscosity with $\eta_0=1.6\times10^{-3}$ in normalized units (Case~5), and with ten times enhanced compressive viscosity coefficient  $\eta_0=1.6\times10^{-2}$ (Case~6)  in Figure~\ref{vdbdnT_T7Mk:fig}. It is evident  in Figure~\ref{vdbdnT_T7Mk:fig}a that the velocity component along the loop at the apex ($V_x$) shows oscillatory evolution, with initial peak due to the arrival of the pulse at the loop apex and shorter period of $70\tau_A$ compared to the cooler loop cases. Using best fit exponential decay to the slow magnetosonic wave velocity component  $V_x$ at the apex as shown in Figure~\ref{fig:vxfit}, we find that the decay rate in the inviscid Case 4 is $t_d=169\tau_A$, in Case~5 with $\eta_0=1.6\times10^{-3}$ the damping time is $t_d=110\tau_A$, while in Case~6 with $\eta_0=1.6\times10^{-2}$ the damping time is $t_d=54\tau_A$. Here as well the magnetic field $\Delta B_x/B_0$ and density oscillations at the apex show about half of the period of $\sim35\tau_A$, since they are due to the nonlinear fast magnetosonic wave, while $\delta B_z/B_0$ is about half the amplitude of $V_x$ and phase shifted by half wave period suggesting vertical kink mode oscillations. The small amplitude magnetic oscillation (Figure~\ref{vdbdnT_T7Mk:fig}b) at the apex show the same period as the density oscillation, due to the variation of the total pressure perturbation resulting in nonlinear coupling between the slow and fast magnetosonic waves. We note, that when a standing slow mode wave is formed in the loop, the apex of the loop has a node in density perturbations.  All variable oscillations show considerable dependence on  $\eta_0$, demonstrating the coupling of the pulse with nonlinear fast magnetosonic oscillations. The evolution of the density perturbation $\Delta\rho$ and temperature perturbation $\Delta T$ is shown in Figure~\ref{vdbdnT_Tr3:fig}c. The temperature is also affected by the change in the background equilibrium, as evident from the non-oscillatory decrease of the temperature at the apex (see, Figure~\ref{vart_case5:fig}g), but with a smaller relative magnitude than in Cases~2-4 (note the difference of the normalizing temperature $T_0$ indicated in Table~\ref{param:tab}). It is evident that the compressive viscosity in a hot loop ($T=10.5Mk$) has stabilizing effect on the background non-equilibrium, since the viscous force damps flows along the field, and hence decreases the associated changes in the loop structure.

%13
The space-time plot of the density and velocity oscillations along the loop due to the flow injection for Cases~ 4-6 are shown in Figure~\ref{xt_T7Mk:fig} for the hot loop with peak temperature $T_{\rm max}$=10.5MK. Case~4 is without viscosity, Case 5 is with classical value of the compressive viscosity coefficient, $\eta_0=1.6\times10^{-3}$, and Case~6 with ten times larger than classical viscosity coefficient  $\eta_0=1.6\times10^{-2}$. It is similar to the case when $T_0$=2 MK that the initial disturbances show the upward wave front propagates at a speed faster than the downward wave front. With the linear best fits, we measured their propagating speeds $V_{p1}=775\pm10$ km~s$^{-1}$ and $V_{p2}=391\pm10$ km~s$^{-1}$ for Case 4, $V_{p1}=705\pm36$ km~s$^{-1}$ and $V_{p2}=357\pm20$ km~s$^{-1}$ for Case 5, and $V_{p1}=731\pm37$ km~s$^{-1}$ and $V_{p2}=373\pm20$ km~s$^{-1}$ for Case 6. Compared to the slow magnetosonic speed $C_s$=428 km~s$^{-1}$ for peak temperature in the loop $T_{\rm max}$=10.5 MK, we obtained the mean upward propagation speed $\bar{V}_{p1}=(1.72\pm0.08)C_s$, and the mean downward propagation speed $\bar{V}_{p2}=(0.87\pm0.04)C_s$. The downward wave speed is about half the upward wave speed. The strong effect of the compressive viscosity damping of the wave amplitude is evident in the classical as well as the enhanced viscosity cases compare to the inviscid model in the density (left panels) and velocity (right panels) perturbations. The effects of the gradual loop background variability is less important on these timescales, compared to the lower temperature loop in Cases~1-3. We have also tested the effect of injected velocity magnitude on the excitation of the slow magnetosonic waves and on the nonlinear effects. In Case~7 velocity amplitude was reduced by a factor of 10 with respect to Case~4 of the 3D MHD model AR (all other parameters were identical). As expected, the smaller injection amplitude $V_{i,0}=0.01$, produces smaller amplitude slow magnetosonic waves, that appear more linear (i.e., sinusoidal) with even smaller amplitude nonlinear fast magnetosonic waves  compared to the larger amplitude waves produced with $V_{i,0}=0.1$ in Case~4. Comparing the amplitudes of the first peak of the oscillations at the loop apex following the flow injection with $V_{i,0}=0.01$ we find that the slow magnetosonic waves velocity amplitude $V_x$ has decreased by a factor of $\sim$14,  the vertical magnetic oscillations amplitude in $B_z$ decreased by a factor of $\sim18$, while the magnetic oscillations amplitude in $B_x$ has decreased by factor of $\sim35$ compared to Case~4 with $V_{i,0}=0.1$, i.e., exhibiting nonlinear dependence of the amplitudes on $V_{i,0}$.

 %We compare the 3D MHD results to the linear analytic solution of the inviscid 1D model in uniform background along a straight magnetic field of length $L$ to the 3D model of the loop with the same loop length, with slow magnetosonic speed nearly uniform along the loop axis in our model, due to the polytropic index $\gamma=1.05$ close to isothermal value ($\gamma=1$). Thus, for the cases with $T=10.5$MK loop axis temperature for the fundamental standing mode we use $C_s=2L/P$ from the analytic 1D model \citep[e.g.,][]{ER83} with $V_A>>C_s$. From the 3D MHD model solution we obtain the waves period of $\sim70\tau_A=707$ s with the loop length of $L=149$ Mm leading to $C_s=2\times149000/707=422$ km s$^{-1}$, in excellent agreement of these parameters from the analytic 1D model. The analytic 1D solution of the inviscid slow magnetosonic wave equation is undamped, and sinusoidal \citep[see,][]{ER83}, while the damping due to leakage and viscosity as well as the nonlinear features are apparent in our 3D MHD model solutions.
 
 We compare the 3D MHD results of the propagation and damping of slow magnetosonic waves to 1D numerical model along the coronal loop, approximated with a uniform background density and temperature using average values along the loop in Cases 4-7 with length $L=149$ Mm. The background magnetic field is taken to be uniform along the loop, and the flow was injected with small amplitude $V_0=0.00145V_A$ (i.e., linear regime) at the boundary in the 1D model. We find that for the cases with $T_0=10.5$MK  the fundamental mode period is $P\sim70\tau_A=707$ s in the 1D model, in excellent agreement with the period obtained with the 3D model, as well as with the fundamental standing mode period obtained from the analytical 1D model using the expression $C_s=2L/P$ \citep[e.g.,][]{ER83} and $V_A>>C_s$. 
 
 Further, we compared the viscous damping times of the slow mode waves computed with the 3D MHD model to the 1D model computations, and found that the viscous dissipation times of the slow magnetosonic waves obtained with the 1D model were different by approximately a factor of two compared to the 3D MHD dissipation times. For classical viscosity coefficient for this temperature the damping was $t_d=210\tau_A$  in the 1D model compared to $t_d=109\tau_A$ in 3D MHD model; for enhanced viscosity by a factor of ten of the classical value  the damping time in 1D model was $t_d=25\tau_A$ compared to $t_d=54\tau_A$ obtained in 3D MHD model. For the inviscid case the slow waves were practically undamped in 1D model, while the damping time was $t_d=169\tau_A$ in the 3D MHD model, consistent with leakage as the main damping mechanism in the inviscid 3D coronal loop case. Thus, while the fundamental mode period is reproduced well in 3D MHD and in 1D models there are considerable difference between the 1D model and nonlinear 3D MHD calculations of the slow mode wave damping times strongly suggesting that 3D MHD modeling is required to accurately account for slow magnetosonic wave viscous damping times in realistic coronal loops.

Figure~\ref{xt_T7Mk:fig} reveals the important role of compressive viscosity in quick formation of the standing mode from initially excited propagating slow magnetosonic pulse. We find that the initial propagating perturbations transition into a standing wave pattern (characterized by nearly opposite-phase oscillations between the two legs) after about four reflections in Case 4 with $\eta_0$=0 (see Figure~\ref{xt_T7Mk:fig}a). The formation time of the standing wave reduces to about two reflections in Case 5 with $\eta0$ taking the classical value (see Figure~\ref{xt_T7Mk:fig}b). While the standing wave forms quickly following only one reflection in Case 6 (see Figure~\ref{xt_T7Mk:fig}c) when $\eta_0$ is enhanced by an order of magnitude compared to the classical value. The formation of a fundamental mode in Case 6 is also evident from the node of the density forming in the middle (apex) of the loop (see Figure~\ref{vdbdnT_T7Mk:fig}c). Despite some effects from wave leakage due to nonlinear coupling with other wave modes, the rapid formation of the standing wave due to enhanced viscosity in the 3D MHD model is consistent with that obtained based on 1D MHD simulations by \citet{Wan18, WO19} and supports the conclusion that quick excitation of the fundamental standing slow mode wave, as observed in some flaring hot loops \citep[e.g.,][]{Wan03b,Wan15}, could be due to anomalous enhancement of compressive viscosity along the magnetic field due to a physical process  that remains to be fully understood.

\section{Discussion and Conclusions} \label{disc:sec}

Recent studies used coronal seismology techniques to analyze slow-mode magnetosonic loop oscillation and damping observed with SDO/AIA, found evidence for suppressed thermal conduction and significantly enhanced compressive viscosity in flaring hot ($T\sim10$ MK) coronal loops. Furthermore, the transport coefficients were estimated based on a parametric study of wave properties using 1D nonlinear MHD loop model in combination with the linear theory. However, neither the past 1D studies (nor 2.5D MHD studies that exclude out-of-plane wave mode coupling) are capable to account fully nor self consistently for the leakage of the slow magnetosonic wave in the coronal part of the loop, nor for the  nonlinear coupling to other wave modes, such as fast magnetosonic and kink oscillation. These couplings could significantly affect hot coronal loops, and impact coronal seismology application.

 While main trapping of the slow magnetosonic waves in low-$\beta$ plasma is by the magnetic field, where in the limit of infinite $B_0$ and $\beta\rightarrow 0$ the slow magnetosonic are confined to travel strictly along the magnetic field lines \citep[see the review][]{NK20}. However, in curved, finite magnetic field, the leakage could play an important role as found in past studies \citep{Sel07,SO09,Ofm12}. In particular, the importance of the slow magnetosonic wave leakage and nonlinear coupling was  demonstrated in a 3D MHD study by \citet{Ofm12} of hot coronal loops in a model AR. While the enhanced temperature cross-field structure of the coronal loops in the present model the trapping of slow magnetosonic waves by forming a leaky waveguide inside the loop (since $T_r>1$, see the discussion of leaky waveguides in \citet{Dav85}), evident by comparing the damping rate in Figure~\ref{vdbdnT_Tr3:fig} to the damping rate \citet{Ofm12} Figure~8), the leakage is still significant in causing the slow-mode damping as evident from the present  inviscid results and with classical values of the compressive viscosity coefficient. However, the enhanced compressive viscosity suppresses the nonlinearity effects by quickly reducing the wave amplitude, and hence the effects of leakage become relatively (to the inviscid or classical viscosity cases) less important in the damping, and in these cases the damping is dominated by the viscous dissipation of the slow magnetosonic waves.

Using 3D MHD model of a bipolar AR initialized with background hot and dense coronal loop we demonstrate the excitation and dissipation of slow magnetosonic waves in realistic coronal loop geometry,  with compressive viscosity along the magnetic field. We investigate the effects of classical and enhanced compressive viscosity coefficients on the propagation and dissipation of the waves in hot coronal loops and compare the results to the inviscid 3D MHD model. We find that in the case of hot loop with peak temperature of $T_{\rm max}$=6 MK the effects of classical as well as enhanced compressive viscosity are small on the loop oscillation dissipation, and the main effects are leakage  and nonlinear coupling to fast magnetosonic (or kink) waves, in agreement with our previous inviscid model 3D MHD AR studies. However, we find that for hotter loop with peak temperature $T_{\rm max}$=10.5 MK, both classical compressive viscosity (that is significantly larger than in 6MK loops due to the viscous coefficient scaling as $\eta_0\propto{T}^{2.5}$), as well as enhanced compressive viscosity play important roles in the damping of the slow magnetosonic waves in the model coronal loops. In addition, the faster $C_s$ (i.e., higher frequency) wave in the hotter loop is subject to increased effect of viscous dissipation compared to the lower temperature (lower $C_s$) wave dissipation, due to the longer path traveled by the wave in the viscous medium in the same time interval. Comparison of the viscous damping times obtained from the 3D MHD model to more approximate 1D calculations show significant differences (factor of $\sim2$) in the results.

We find from our model that the enhance viscosity can facilitate the rapid formation of a standing slow magnetosonic wave in the hot flaring coronal loop by damping quickly the higher harmonics, resulting in fundamental mode oscillation, in agreement with the observations reviewed in Section~\ref{obs:sec}.  Moreover, the 3D MHD model demonstrated that the enhanced compressive viscosity leads to extremely rapid damping within a couple of oscillation periods as seen in observations. The decreased amplitude of the wave due to the viscous damping dissipates the nonlinearity effects, such as wave steepening, and leads quickly to the linear regime of small amplitude oscillations. Our model also demonstrates that the changes in the initial background hot loop  structure (seen in SDO/AIA in hot flaring loops by, e.g., \citet{Wan18}) may increase the leakage rate of the slow magnetosonic waves. The present more realistic 3D MHD model then previous studies of AR with a hot and dense loop that considers the effects of compressive viscosity on the dissipation of slow magnetosonic waves allows improved determination of dissipation coefficient in hot flaring coronal loops, important for coronal seismology and for the understanding of coronal heating processes.

%\begin{acknowledgments}
The authors acknowledge support by NASA grant 80NSSC18K1131 and by NASA Cooperative Agreement  80NSSC21M0180 to The Catholic University of America. T.W. also acknowledges support by the NASA grant 80NSSC18K0668. Resources supporting this work were provided by the NASA High-End Computing (HEC) Program
through the NASA Advanced Supercomputing (NAS) Division at Ames Research Center.
%\end{acknowledgments}


\vspace{5mm}
\facilities{SDO (AIA)}

\end{document}